\documentclass[aps,pre,twocolumn,showkeys,preprintnumbers,floatfix]{revtex4-1}

\usepackage{dynlearn}

\def\deriv#1 #2{\frac{\text{d}}{\text{d}#2} #1}

\def\npderiv#1 #2 #3{\frac{\partial^#1}{\partial #3^#1} #2}
\def\fpderiv#1 #2{\frac{\partial #1}{\partial #2} }
\def\fnpderiv#1 #2 #3{\frac{\partial^#1 #2}{\partial #3^#1}}
\def\fnderiv#1 #2 #3{\frac{\dd^#1 #2}{\dd #3^#1}}

\def\comm#1,#2.{\left[ #1,#2\right]}

\def\dd{\text{d}}

\def \df#1{\hat{#1}}
\def \dl#1{#1}

\newcommand{\p}{\varphi}
\newcommand{\px}{\varphi_x}
\newcommand{\pdc}{\varphi_{\text{dc}}}
\newcommand{\pxdc}{\varphi_{\text{xdc}}}
\newcommand{\kB}{k_\text{B}}
\newcommand{\VS}{V^{\text{store}}}
\newcommand{\VC}{V^{\text{comp}}}
\newcommand{\pxS}{\px^{\text{store}}}
\newcommand{\pxC}{\px^{\text{comp}}}
\newcommand{\pxdcS}{\pxdc^{\text{store}}}
\newcommand{\pxdcC}{\pxdc^{\text{comp}}}

\graphicspath{{./img/}}

\theoremstyle{plain}    

\begin{document} 

\def\ourTitle{Gigahertz Sub-Landauer Momentum Computing}

\def\ourAbstract{We introduce a fast and highly-efficient physically-realizable
bit swap. Employing readily available and scalable Josephson junction
microtechnology, the design implements the recently introduced paradigm of
momentum computing. Its nanosecond speeds and sub-Landauer thermodynamic
efficiency arise from dynamically storing memory in momentum degrees of
freedom. As such, during the swap, the microstate distribution is never near
equilibrium and the memory-state dynamics fall far outside of stochastic
thermodynamics that assumes detailed-balanced Markovian dynamics. The device
implements a bit-swap operation---a fundamental operation necessary to build
reversible universal computing. Extensive, physically-calibrated simulations
demonstrate that device performance is robust and that momentum computing can
support thermodynamically-efficient, high-speed, large-scale general-purpose
computing that circumvents Landauer's bound.
}

\def\ourKeywords{rate equations, stochastic process, information processing, logical circuits,
entropy production, reversibility
}

\hypersetup{
  pdfauthor={James P. Crutchfield},
  pdftitle={\ourTitle},
  pdfsubject={\ourAbstract},
  pdfkeywords={\ourKeywords},
  pdfproducer={},
  pdfcreator={}
}

\author{Kyle J. Ray}
\email{kjray@ucdavis.edu}
\affiliation{Complexity Sciences Center and Physics Department,
University of California at Davis, One Shields Avenue, Davis, CA 95616}

\author{James P. Crutchfield}
\email{chaos@ucdavis.edu}
\affiliation{Complexity Sciences Center and Physics Department,
University of California at Davis, One Shields Avenue, Davis, CA 95616}

\title{\ourTitle}

\begin{abstract}
\ourAbstract
\end{abstract}


\date{\today}

\preprint{arxiv.org:2202.07122 [cond-mat.stat-mech]}

\maketitle

\section{Introduction}

Ever since the first exorcism of Maxwell's demon \cite{Szil29a}, determining
how much energetic input a particular computation requires has been a
broadly-appreciated theoretical question. In the current century, however, the
question has taken on a markedly practical bent; a familiar example is the
evolution of Moore's Law from initially provocative speculations decades ago to
now addressing material, thermodynamic, and fabrication restrictions
\cite{moore1964, moore1998, moore2006, hutcheson1993, hutcheson1996}.
Transistor-based microprocessing presents fundamental scaling challenges that
strictly limit potential directions for future optimization, and these
challenges are no longer speculative. Clock speed, to take one example, has
been essentially capped for two decades due to energy dissipation at high rates
\cite{gelsinger1989,waldrop2016}. By some measures, Moore's law is already
dead---as integrated circuit manufacturers go vertical, rather than face the
expense of creating smaller transistors for 2D circuits that yield only
marginal gains \cite{ball2022, vinet2011, courtland2021}.

Given predicted explosive growth in societal demands for information processing
and that digital microelectronics is now approaching the physical limits of
available architectures \cite{ITRS20a}, exploring alternative computing
paradigms is not only prudent but necessary. One alluring vision for the future
involves hybrid devices, composed of a suite of computing
modules---classical/quantum, digital/analog, deterministic/thermal---each with
its own architecture and function that operate in concert. A hybrid architecture
allows dynamically harnessing the processing node best suited for the task at
hand. The underlying insight is that a computing device's physical substrate should
match its desired processing function \cite{Feyn82a}. In keeping with this,
momentum computing demonstrated that low dissipation operations do not require
quasi-static operation \cite{ray2021}. That is, energy-efficient computation
can be fast in a low-dissipation device.

Reference \cite{ray2021} introduced a design framework and theory for an
arbitrarily low-cost, high-speed bit swap, a logically-reversible gate (the only known logical framework with no nontrivial
lower bound on its dissipation \cite{ITRS20a, frank2005, bhattacharya2021}.)
It demonstrated that a universal
reversible gate---a Fredkin gate \cite{toffoli1980, Fred82a}---can be built by
coupling three such devices together. However, any particular
physically-instantiated implementation will come with its own restrictions and considerations that are likely to disallow
performing the swap exactly as theorized. And so, an implementation
linked to a particular substrate must be built and analyzed in its own right.

We present a physically-realizable device and control protocols that implement a
bit swap gate that operates in the sub-$\kB T$ energy regime using
superconducting Josephson junctions (JJs)---a well-known and scalable
microtechnology. We recently used this device to measure the thermodynamic
performance of bit erasure \cite{saira2020,Wims19a}. That extensive
experimental effort demonstrated in practical terms that the device proposed
here is realizable with today's microfabrication technologies and allows for
detailed studies of thermodynamic costs. And so, the device's design and
control protocol open up exploring the energy scales of highly
energy-efficient, high-speed, general-purpose computing.

\paragraph*{The Landauer}
While there are many different quantities one might wish to optimize, the perspective here sets the goal as minimizing
the net work invested $W$  when performing logical operations. It is well known
that the most pressing physical limits on modern computation are power
constraints \cite{frank2002}, thus the measure is well suited to diagnose the
problems with current devices as well as potential strengths of new ones.

For over half a century now \emph{Landauer's Principle} has exerted a major
impact on the contemporary approach to thermodynamic costs of information
processing \cite{Land61a,Benn82}. Its lower bound of $\kB T \ln 2$ energy
dissipated per bit erased has served as standard candle for energy use in
physical information processing. To aid comparing other computing
paradigms and protocols, we refer to this temperature-dependent
information-processing energy scale as a \emph{Landauer}: approximately a few
zeptojoules at room temperature, and a few hundredths of a zeptojoule at
liquid He temperatures. See Appendix \ref{sm:Landauer} for further comparisons.
 
To appreciate the potential benefits of momentum computing operating at
sub-Landauer energies we ask where contemporary computing is on the energy
scale. Consider recent stochastic thermodynamic analyses of single-electron
transistor logic gates \cite{gao2021,freitas2021}---analogs to conventional
CMOS technology. The upshot is that these technologies currently operate
between $10^3$ and $10^4$ Landauers. More to the point, devices using
CMOS-based technology will only ever be able to operate accurately above
$\approx 10^2$ Landauers \cite{frank2005,ITRS20a}. In short, momentum computing
promises substantial improvements in efficiency with no compromise in speed.

\paragraph*{Outline} Here we provide a brief overview of each section and appendix in the text. Section \ref{sec:bitswap} explains the importance of bit-swap operations and
summarizes the protocol presented in Ref. \cite{ray2021}. Section
\ref{sec:physswap} introduces the physical substrate, highlights why it is a
good candidate, and addresses design restrictions. Section
\ref{sec:performance} reports quantitative results on device performance as
measured through detailed simulations of the microscopic degrees of freedom.
Section \ref{sec:litreview} compares them to related results, both contemporary
and foundational. Section \ref{sec:conclusion} concludes, summarizing the
results and briefly outlining future directions and challenges for scaling up
to general-purpose computing.

Appendices include details necessary to understand the process by which the
parameter space of control protocols was restricted and local work minima were
found in simulation. Additionally, they also provide expository information
that the interested reader might find relevant. In particular, Appendix
\ref{sm:Landauer} discusses the temperature dependent energy scale, the
``Landauer''. Appendix \ref{sm:LimitStochThermo} outlines key physical
differences between continuous-time Markov chains and hidden Markov chains.
Appendix \ref{sm:DimensionlessEoM} presents the equations of motion of the
bit-swap Josephson junction circuit in their dimensional form and their
transformation to simulation-appropriate dimensionless equations.
Appendix \ref{sm:PotentialSimplifications} details the process of algebraically
eliminating large swaths of protocol parameter space. And, finally, Appendix
\ref{sm:SearchMinWork} discusses the algorithmic details of the simulations.

\section{Bit Swap}
\label{sec:bitswap}

The Landauer cost stood as a reference for so long since bit erasure is the
dominant source of unavoidable dissipation when implementing universal
computing with transistor logic gates. It is the elementary binary computation
that most changes the Shannon entropy of the distribution over memory states.
In this way, one sees $\kB T\ln 2$ not just as the cost of erasure, but as the
cost of the maximally dissipative elementary operation on which conventional
computing relies. And so, the Landauer naturally sets the energy scale for
conventional computing.

Taking inspiration from Landauer's pioneering work, we investigate
the cost of the most expensive operation necessary to physically implement
universal momentum computing: a bit swap. The ideal bit swap has no error, but
in the thermodynamic setting one is also interested in an implementation's
fidelity. And so, we write a swap with error rates $\epsilon_0$ and
$\epsilon_1$ as a stochastic mapping between memory states $m \in \{0,1\}$ from time $0$ to
time $\tau$:
\begin{align*}
P_\epsilon(m_\tau | m_0)
  = \left[\begin{array}{cc}
  \epsilon_0 & 1-\epsilon_0 \\
  1-\epsilon_1 & \epsilon_1
  \end{array}\right]
  ~.
\end{align*}

The bit swap's dominance in the cost of universal momentum computing can
be appreciated by considering the input-output mapping of the Fredkin gate---a
$3$-bit universal gate with memory states $m_x m_y m_z$, $m_i \in \{0,1\}$. All
inputs are preserved except for the exchange $101 \leftrightarrow 110$. We can
decompose the informational state space into two regions. If $m_x = 0$, the
operation is simply an identity, which trivially is costless. If $m_x = 1$ and $m_y = m_z$, we once again have an identity. Thus, it is only the subspace of $m_x=1$, where $m_y \neq m_z$ that a swap must take place. Reference  \cite{ray2021} provides explicit potentials that impose effectively 1D swap potentials on a full 3-bit state space in order to implement the Fredkin gate, demonstrating that only 1D swap operations need contribute to the
operation's thermodynamic cost.

\subsection{Momentum Computing Realization}

Storing information in a one-dimensional state space, it is not clear how to
operate a thermodynamically-efficient bit swap with high accuracy. (In this, we
recall the conventional interpretation of efficient to mean quasistatic or
constantly-thermalizing Markovian dynamics \cite{esposito2012, seif2019}.) At time $t$ in the operation, the
 distribution of initial conditions corresponding to $m(t = 0)=0$ must overlap with
that corresponding to $m(t = 0)=1$. And, from that point forward it is
impossible to selectively separate them based on their initial positions.
Information, and so reversibility, is lost.

Consider, instead, a computation that happens
faster than the equilibration timescale of the physical substrate and its
thermal environment. In this regime, a particle's instantaneous momentum can be
commandeered to carry useful information about its future
behavior. Our protocol operates on this timescale, using the full phase space of the
underlying system's degrees of freedom to transiently store information in
their momenta. Due to this, the instantaneous microstate distribution is
necessarily far from equilibrium during the computation. Moreover, the
coarse-grained memory-state dynamics during the swap are not Markovian; despite
both the net transformation over the memory states and the microscopic phase
space dynamics being Markovian. Nonetheless, the system operates orders of
magnitude more efficiently than current CMOS but, competing with CMOS, the
dynamics evolve nonadiabatically in finite time---on nanosecond timescales for
our physical implementation below.

In this way, momentum computing offers up device designs and protocols that
accomplish information processing that is at once fast, efficient, and low
error. There is a trade-off---a loss of Markovianity in the memory-state dynamics.
That noted, the dynamics of the memory states are faithfully described by
continuous-time \emph{hidden} Markov chains (CTHMCs) \cite{bech2015,
strasberg2016, ara2016}, rather than the continuous-time Markov chains (CTMCs)
that are common in stochastic thermodynamics \cite{esposito2012, seif2019}. See
Appendix. \ref{sm:LimitStochThermo} for a brief review.

\subsection{Idealized Protocol}
\label{sec:exact_protocol}

Reference \cite{ray2021} describes a perfectly-efficient protocol for
implementing a swap in finite time. The operation is straightforward. We begin
with an ensemble of particles subject to a storage
potential. The potential energy landscape $V^\text{\text{store}}(x)$ must
contain at least two potential minima---positioned, say, at $x = \pm
x_0$---with an associated energy barrier equal to $ \text{max}\{
V^{\text{store}}(x), x\in(-x_0,x_0) \} - V^{\text{store}}(x_0)$. During
storage, a particle's environment is a thermal bath at temperature $T$. As the
height of the potential energy barrier rises relative to the bath energy scale
$\kB T$, the probability that the particle transitions between left ($x<0$) and
right ($x\geq0$) decreases exponentially. In this way, if we assign the left
half of the position space to memory state $0$ and the right half to memory
state $1$, the energy landscape is capable of metastably storing a bit $m \in
\{0,1\}$.

At the protocol's beginning, we instantaneously apply a new potential energy
landscape $V^{\text{comp}} \equiv k x^2 / 2$. The system is then temporarily
isolated from its thermal environment, resulting in the particles undergoing a
simple harmonic oscillation. Waiting a time $\tau$ until the oscillation is
only half completed, the potential is returned to $V^{\text{store}}$. The
initial conditions---for which $x_0<0$ $(x_0>0)$---have then been mapped to
$x_\tau >0$ $(x_\tau<0)$, achieving the desired swap computation. If
$V^{\text{store}}$ is an even function of $x$, the computation requires zero
invested work as well. This follows since the harmonic motion created a mirror
image to the original distribution and the energy imparted to the system at
$t=0$ is completely offset by the energy extracted from the system turning off
$V^{\text{comp}}$ at $t=\tau$.

\section{Physical Instantiation}
\label{sec:physswap}

Due to its conceptual simplicity the protocol does not require any particular
physical substrate. That said, the practical feasibility of performing such a
computation must be addressed. One obvious point of practical concern is
assuming the system can be isolated from its thermal environment during the
computation. However, total isolation is not necessary. If $\tau \ll
\tau_R$---the relaxation timescale associated with the energy flux rate between
the system and its thermal bath---then the device performs close to the ideal
case of zero coupling.

As proof of concept, Ref. \cite{ray2021}'s simulations showed that this class
of protocol is robust: thermodynamic performance persists in the presence of
imperfect isolation from the thermal environment, albeit at an energetic cost.
Thus, a system that obeys significantly-underdamped Langevin dynamics is an
ideal candidate as the physical substrate for bit swap.

\begin{figure}[t]
\centering
\includegraphics[width=\columnwidth]{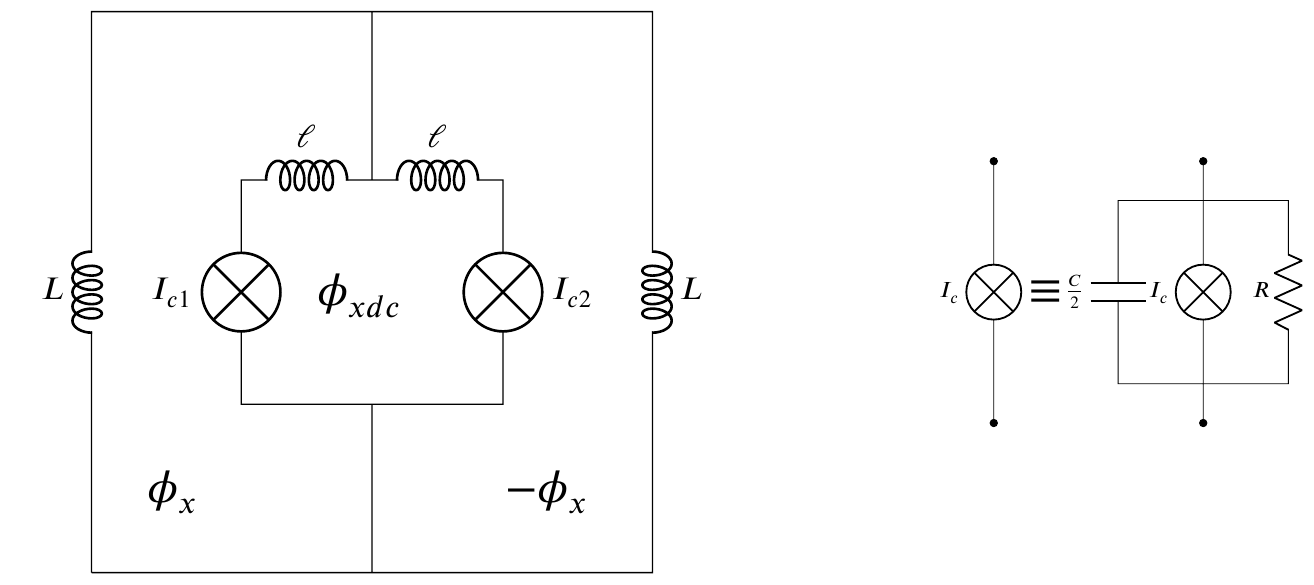}
\caption{Gradiometric flux logic cell: The superconducting current has two
	important flow modes. One circulation around the inner loop---a DC SQUID.
	And, the other, a flow through the Josephson junctions in the inner loop
	and around the outer conductor pickup loops---an AC SQUID
	\cite{han1992theory}. This is the origin of the variable subscripts to
	distinguish $\p$ from $\pdc$ and $\px$ from $\pxdc$.
  }
\label{fig:circuit}
\end{figure}

We analyze in detail one physical instantiation---a \emph{gradiometric flux
logic cell} (Fig. \ref{fig:circuit}), a mature technology for information
processing. With suitable scale definitions, the effective degrees of
freedom---Josephson phase sum  $\p$ and difference $\pdc$---follow a
dimensionless Langevin equation \cite{barone1982, han1992theory,
han1992experiment, rouse1995, saira2020, Wims19a}:
\begin{align}
dv' = -\lambda v' dt' - \theta \partial_{x'} U' + \eta r(t) \sqrt{2dt'}
  ~,
\end{align} 
where $x' \equiv (\p,\pdc)$ and $v' \equiv (\dot{\p},\dot{\pdc})$ are vector representations of the dynamical coordinates. Enacting a control protocol on this system involves changing the parameters of the potential over time:
\begin{align}
U'(t') & = U/U_0 \\
  & = (\p-\px(t'))^2/2 + \gamma (\pdc-\pxdc(t'))^2/2 \nonumber \\ 
  & \qquad + \beta \cos \p \cos (\pdc/2) - \delta\beta \sin \p \sin (\pdc/2) \nonumber
~.
\end{align}

\begin{figure}[t]
\centering
\includegraphics[width=\columnwidth]{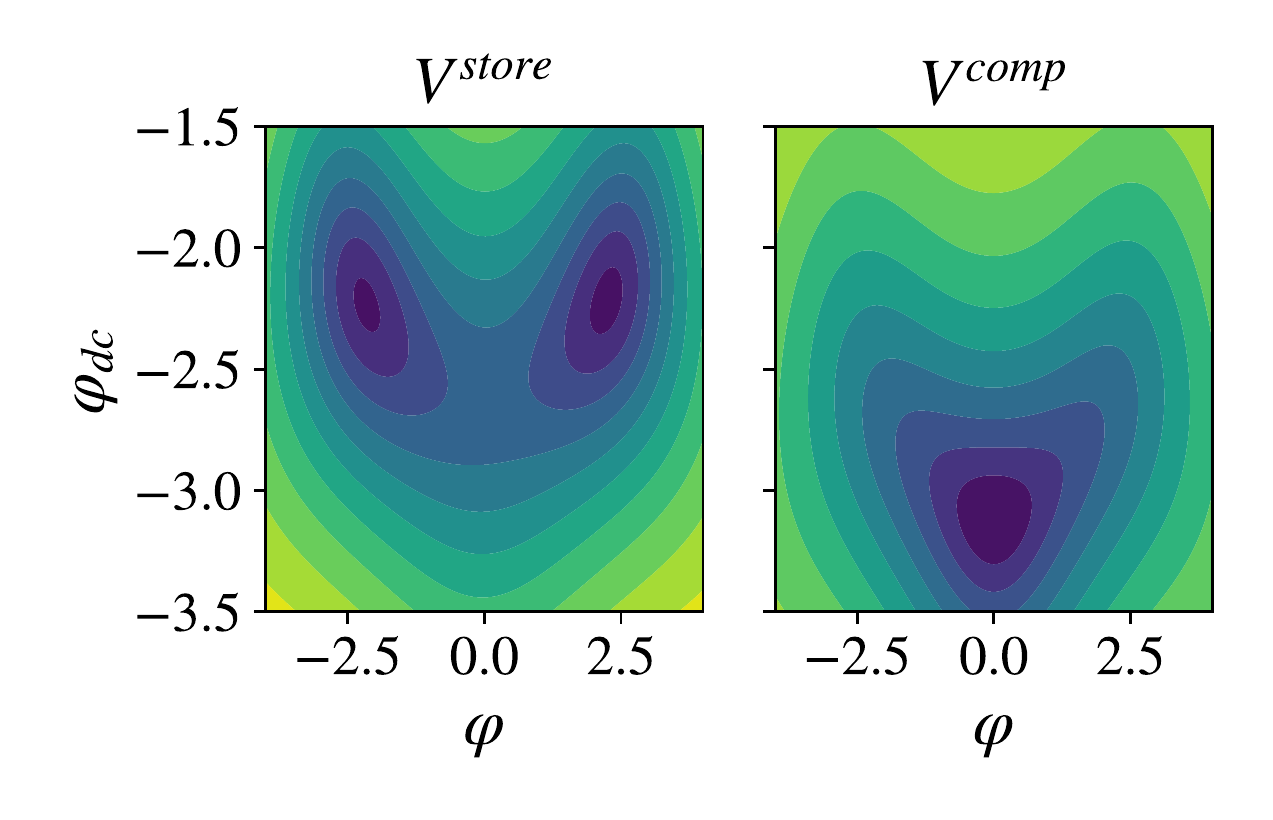}
\caption{(Left) $V^{\text{store}}$, the bistable storage potential.
	(Right) $V^{\text{comp}}$, the ``banana-harmonic'' potential. These potential energy profiles serve as qualitative pictures to represent prototypical computational and storage potentials, and do not represent any particularly favorable parameter set. 
	}
\label{mainfig:potentials}
\end{figure}

The relationships between the
circuit parameters and the parameters in the effective potential $U'$ are as follows. $\p
= (\p_1+\p_2)/2 -\pi$ and $\pdc = (\p_2-\p_1)$, where $\varphi_1$ and
$\varphi_2$ are the phases across the two Josephson elements; $\px = 2\pi
\phi_x/\Phi_0 - \pi$ and $\pxdc = 2\pi \phi_{xdc}/\Phi_0$, where $\Phi_0$ is
the magnetic flux quantum and $(\phi_x, \phi_{xdc})$ are external magnetic
fluxes applied to the circuit; $U_0 = \left(\Phi_0 / 2\pi\right)^2 / L$,
$\gamma = L / 2\ell$, $\beta = I_+ 2\pi L / \Phi_0$, and $\delta\beta = I_- 2\pi
L / \Phi_0$, where $L$ and $2\ell$ are geometric inductances; and $I_{\pm}
\equiv I_{c1} \pm I_{c2}$ are the sum and difference of the critical currents
of the two Josephson junctions. All parameters are real and it is assumed that $\gamma > \beta >1 \gg \delta\beta$.

Some particularly important parameters of $U'$ are $\px$ and $\pxdc$, which control the potential's shape by
where the the dynamical variables $\p$ and $\pdc$
localize in equilibrium, and $\gamma$, which controls how quickly $\pdc$
localizes to the bottom of the quadratic well centered near $\pdc=\pxdc$. At certain control parameters ($\px, \pxdc$), the effective potential contains only two
minima: one located at $\p<0$ and one at $\p>0$. So, the device is capable of
metastably storing a bit, as described above. In point of fact, the logic cell
has been often used as a double well in $\p$ with a controllable tilt and
barrier height \cite{han1992theory, rouse1995, saira2020}.

The Langevin equation's coupling constants, $\lambda$ and $\eta$, determine the
rate of energy flow between the system and its thermal environment and the. They depend
on the parameters $L$, $R$, and $C$. In the regimes at which one typically
finds $L$, $C$, and $R$ and with temperatures around $1$ K, the system is very
underdamped; ring-down times are $\mathcal{O}(10^3)$ oscillations about the
local minima. (Notably, the device thermalizes at a rate proportional to
$R^{-1}$. A tunable $R$ allows the device to transition from the underdamped to
overdamped regime, allowing for rapid thermalization, if desired.) Finally, $\theta$ is a dimensionless factor that depends on the relative inertia of the two degrees of freedom, it depends on the circuit architecture. Appendix \ref{sm:DimensionlessEoM} gives the equations of motion and thorough definitions of all parameters and variables in terms of dimensional quantities.

\subsection{Realistic Protocol}

With the device's physical substrate set, we now show how to design
energy-efficient bit-swap control protocols. There are four parameters that
depend primarily on device fabrication: $I_{c1}$, $I_{c2}$, $R$, and $C$. Two
that depend on the circuit design: $L$ and $\ell$. And, four that allow
external control: $\px$, $\pxdc$, $T$ (the environmental temperature), and
$\tau$ (the computation time). Without additional circuit complexities to allow
tunable $L$, $R$, and $C$, we assume that once a device is made, any given
protocol can only manipulate $\px$, $\pxdc$, $T$, and $\tau$. A central
assumption is that computation happens on a timescale over which the thermal
environment has minimal effect on the dynamics, so the primary controls are
$\px$, $\pxdc$, and $\tau$. $\px$ is associated with asymmetry in the
informational subspace, and will only take a nonzero value to help offset
asymmetry from the $\delta\beta$ term in $U'$. Thus, $\pxdc$ primarily controls
the difference between $V^{\text{comp}}$ and $V^{\text{store}}$, while $\tau$
governs how long we subject the system to $V^{\text{comp}}$.

$V^{\text{store}}$ must be chosen to operate the device in a parameter regime
admitting two minima on either side of $\p=0$ as in Fig.
\ref{mainfig:potentials}. They must also be sufficiently separated so that they
are distinct memory states when immersed in an environment of temperature $T$.

In the ideal case, $V^{\text{comp}}$ is a quadratic well with an oscillation
period $\tau = \pi \sqrt{m/k}$. However, $U$ will never give an exact quadratic
well unless $\beta = \delta\beta = 0$. So, a suitable replacement is necessary.
The closest approximate is at the relatively obvious choice $\pxdc = -2\pi$. In
this case, the minima of both the quadratic and the periodic part of the
potential lie on top of each other and the potential is well approximated by a
quadratic function over most of the relevant position-domain.

However, due to restrictions on $\VS$, transitioning between $V^{\text{store}}$
and $V^{\text{comp}}$ may induce unnecessarily large dissipation since the
oscillations in the $\pdc$ dimension have a large amplitude. (See Appendix
\ref{sm:PotentialSimplifications} for details.) Instead, to dissipate the
minimum energy, the control parameters must balance placing the system as close
as possible to the pitchfork bifurcation where the two wells merge, while still
maintaining dynamics that induce the $\p<0$ and $\p>0$ informational states to
swap places due to an approximately harmonic oscillation. Near this parameter
value, one typically finds a ``banana-harmonic''
potential energy landscape. (See Fig. \ref{mainfig:potentials} for a comparison of the distinct
potential profiles for storage and computation.)

\begin{figure}[t]
\centering
\includegraphics[width=\columnwidth]{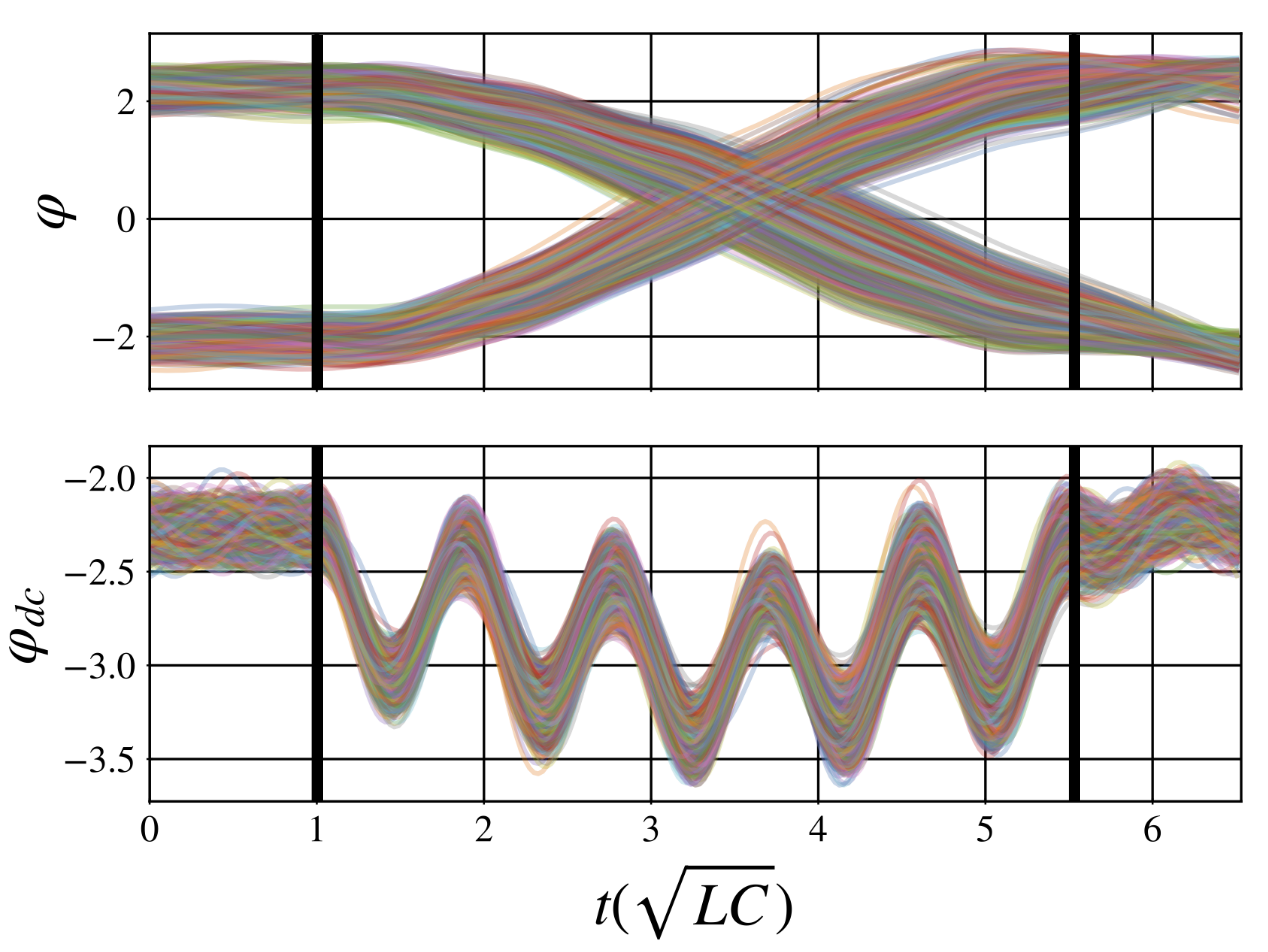}
\caption{A dynamic computation: $1,500$ trajectories from $V^{\text{store}}$'s
	equilibrium distribution in the $\p$ (top) and $\pdc$ (bottom) dimensions.
	$V^{\text{comp}}$ is applied at  $t\in(1,1+\tau)$, denoted by heavy black
	lines. $\pdc$ oscillations are several times faster than the others, as
	expected when $\gamma \gg 1$. The work done on the system by the control
	apparatus, $W_0 = \VC(t=1)-\VS(t=1)$, by its intervention at $t=1$ is
	largely offset by the work absorbed into the apparatus by its intervention
	at $t=1+\tau$, $W_\tau = \VS(t=1+\tau)-\VC(t=1+\tau)$, when $\VC$
	re-engages. Visually, we can track this energy flux by the nonequilibrium
	oscillations induced at $t=1$ and the return to a near-equilibrium
	distribution at $t=1+\tau$. Time is measured in units of $\sqrt{LC}$, which
	is $\approx2$ns for the JJ device.
	}
\label{fig:dynamics}
\end{figure}

\subsection{Computation Time}

The final design task determines the computation timescale $\tau$. Under a
perfect harmonic potential, the most energetically efficient $\tau$ is simply
$\pi \sqrt{m/k}$. This ensures that $x(t=0) = -x(t=\tau)$. Since the design has
an additional degree of freedom beyond that necessary---the $\pdc$
dimension---however, we must not only ensure our information-bearing degree of
freedom switches signs, but also ensure that $\pdc(t=0) \approx \pdc(t=\tau)$.
This means that during time $\tau$, the $\p$ variables must undergo $n+1/2$
oscillations and the $\pdc$ variables must undergo an integer number of
complete oscillations. (See Fig. \ref{fig:dynamics}.) Hence, $\tau$ must
satisfy matching conditions for the periods of the oscillations in both $\p$
and $\pdc$ during the computation:
\begin{align*}
\omega \tau &\approx (2n-1)\pi \\
\omega_{dc} \tau &\approx 2n\pi
  ~.
\end{align*}

\begin{figure}[t]
\centering
\includegraphics[width=\columnwidth]{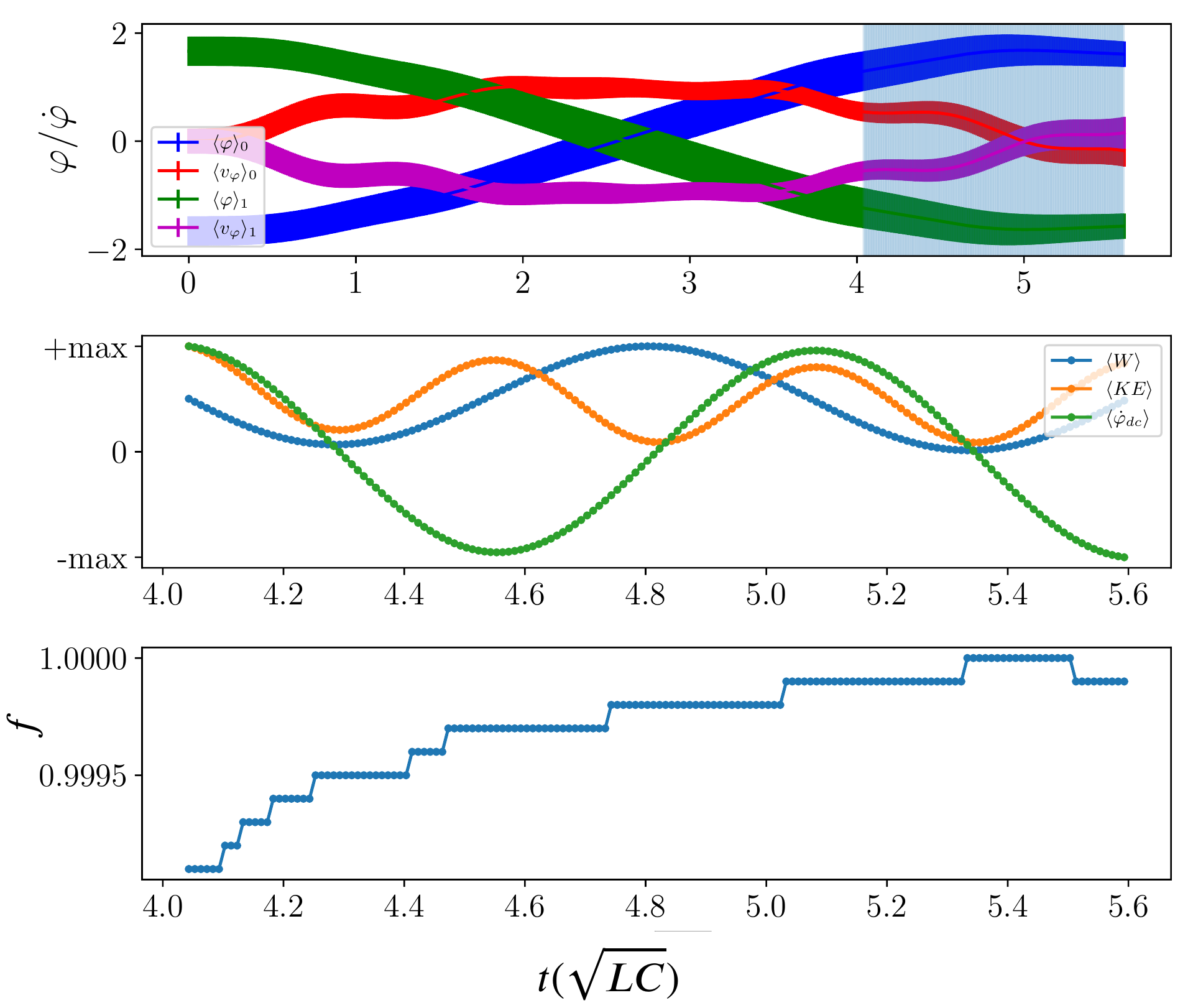}
\caption{Performing a successful and low-cost bit swap: (Top) Ensemble
	averages, conditioned on initial memory state, of the fluxes and their
	conjugate momenta. Line width tracks the distribution's variance. The
	shaded region indicates timescales that are potentially successful swap
	operations. These are probed more closely in the bottom two plots.
	(Middle) Ensemble averaged work, kinetic energy, and conjugate momentum in
	the $\pdc$ coordinate. Note that work minima occur only at whole-integer
	oscillations of the momentum. Each dataset is scaled to its maximum value,
	so that it saturates at $1$. This emphasizes the qualitative relationships
	rather than the quantitative.
	(Bottom) Computational fidelity $f$ of the swap, approaching a perfect swap.
	}
\label{fig:tau_sweep}
\end{figure}

Figure \ref{fig:tau_sweep} showcases this by displaying the behavior observed during simulations near
the ideal timescale. The local work minima coincide with local minima in the average kinetic energy, but not every kinetic energy minimum coincides with a work minimum. While there are kinetic energy minima every half-integer oscillation in $\pdc$,
only integer multiples of $\pdc$ oscillations yield minimum work.

The equations of motion governing the system are stochastic,
dissipative, and nonlinear, so the frequencies of the different oscillations $\omega, \omega_{dc}$
are nontrivial nonlinear stochastic mappings of device parameters, initial positions, and
protocol parameters. They are not easily determined analytically. However, they change smoothly with small changes in the parameters they depends on. Thus, we were able to use an algorithmic approach to find the timescales that yield local minima and explore the regions surrounding them.

\newcommand\xC{\SI{4.0}{\nano\farad}}
\newcommand\xIp{\SI{2.0}{\micro\ampere}}
\newcommand\xR{\SI{371}{\ohm} }

\newcommand\xIm{\SI{7}{\nano\ampere}}
\newcommand\xImb{\SI{35}{\nano\ampere}}
\newcommand\xImbt{\SI{60}{\nano\ampere}}

\newcommand\xL{\SI{1.0}{\nano\henry}}
\newcommand\xT{$0.05 U_0$ }

\newcommand\ntest{50,000}
\newcommand\ntrial{40,000}
\newcommand\dt{$0.005 \sqrt{LC}$}

\subsection{Physically-Calibrated Bit Swap}

We are most interested in the effect of parameters that are least constrained
by fabrication. And so, all simulations assume constant fabrication parameters
with $I_+$, $R$, and $C$ set to $\xIp$, $\xR$, and $\xC$, respectively. To
explore how the $I_-$ asymmetry affects work cost, we simulated protocols with
both a nearly-symmetric device ($I_- = \xIm$) and a moderately-asymmetric
device ($I_- = \xImb$). Given devices with the parameters above, what values of
the other parameters yield protocols with minimum work cost? This involves a
twofold procedure. First, create a circuit architecture by setting $L$ and
$\gamma$, thus fully specifying the device; details in Appendix
\ref{sm:SearchMinWork}. Second, determine the ideal protocols for that
combination of device parameters.

\subsection{Computational Fidelity}

To determine the best successful protocol, we must define what a successful bit swap
is. First, we set a lower bound for the \emph{fidelity} $f$: $f \geq 0.99$. We define $f$ over an ensemble of $N$ independent trials
as: $f = 1 - N_e / N$, with $N_e$ counting the number of failed trials, trials
for which $\text{sign} [\p(t=0)] = \text{sign} [\p(t=\tau)]$. Second, the
distribution over both $\p(t=\tau)$ and $\p(t=0)$ must be bimodal with clear
and separate informational states. The criteria used for this second condition
is:
\begin{align}
\langle \p < 0 \rangle + 3  \sigma_{\p<0} < \langle \p > 0 \rangle - 3  \sigma_{\p>0}
   ~,
 \end{align}
were $\sigma_s$ and $\langle s \rangle$ are standard deviations and means
of $\p$ conditioned on statement $s$ being true.

The final choice concerns the initial distribution from which to sample trial
runs. For this, we used the equilibrium distribution associated with
$V^{\text{store}}$ with the environmental temperature set to satisfy $\kB T =
0.05 U_0$. Here, we ensure fair comparisons between different parameter
settings by fixing a relationship between the potential's energy scale and that
of thermal fluctuations. This resulted in temperatures from $400-1400$ mK,
though it is possible to create superconducting circuits at much higher
temperatures \cite{Yurg00a,Long12a,cybart2015,Revi21a} using alternative
materials.

Sampling initial conditions from a thermal state assumes no special
intervention created the system's initial distribution. We only need wait a
suitably long time to reach it. Moreover, this choice is no more than an
algorithmic way to select a starting distribution. It is not a limitation or
restriction of the protocol. Indeed, if some intervention allowed sampling
initial conditions from a lower-variance distribution, it could be leveraged into even higher
performance.

\section{Performance}
\label{sec:performance}

Appendix \ref{sm:SearchMinWork} lays out the computational strategy used to
find minimal $\langle W \rangle$ implementations among the protocols that satisfy the conditions above. Since the potential is held constant between $t=0$ and $t=\tau$,
work is only done when turning $\VC$ on at $t=0$ and turning it off at
$t=\tau$. The ensemble average work done at $t=0$ is $W_0 \equiv \langle
\VC(\p(0),\pdc(0)) - \VS(\p(0),\pdc(0)) \rangle $ and returning to $\VC$ at
time $\tau$ costs $W_\tau \equiv \big\langle \VS(\p(\tau),\pdc(\tau)) -
\VC(\p(\tau),\pdc(\tau))\big\rangle$. Thus, the mean net work cost is the sum $\langle W \rangle=W_0 + W_\tau$. As we detail shortly, this yielded large regions of parameter
space that implement bit swaps at
sub-Landauer work cost. This result and others demonstrate the notable and
desirable aspects of momentum computing: accuracy, low thermodynamic cost, and
high speed. Let's recount these one by one.

\subsection{Accuracy}

Tradeoffs between a computation's fidelity and its thermodynamic cost are now
familiar---an increase in accuracy comes at the cost of increased $W$ or
computation time \cite{Boyd18a, lahiri2016, zulkowski2014, berut2012,
riechers2020, gammaitoni2011}. These analyses conclude that accuracy generally
raises computation costs.

\begin{figure}[t]
\centering
\includegraphics[width=\columnwidth]{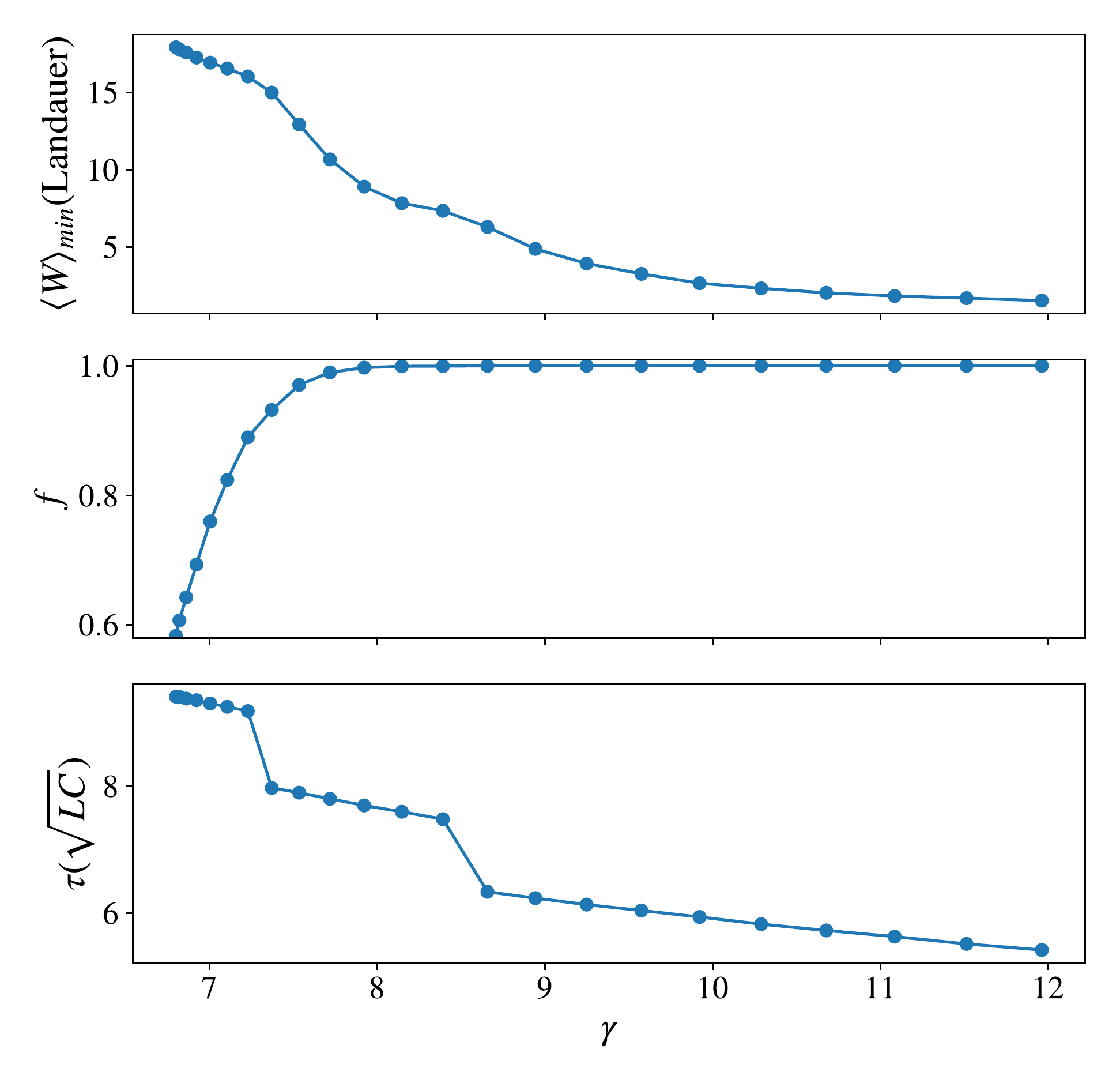}
\caption{Performance of the minimum work protocol as $\gamma$, the ratio of
	device inductances, goes from a region where the computation fails
	($f < 0.99$) to a region of perfect fidelity ($f = 1.0$). Note that in the
	parameter space region in which the computation becomes successful, the
	work costs decrease as the fidelity approaches unity. Finally, $\tau$
	decreases as the work cost minimizes to $\approx 1$ Landauer---showing that
	the work cost does not display $1/\tau$ adiabatic compute-time scaling. The
	parameter $\gamma$ controls the starting parameters for the suite of
	simulations represented by each data point and should not be read as the
	primary independent variable responsible for the behavior. Rather, the
	plots show $\tau$, $f$, and $\langle W \rangle_{min}$ evolving jointly to
	more preferable values.
	}
\label{fig:fid_work}
\end{figure}

Momentum computing does not work this way. In fact, it works in the opposite
way. The low cost of a momentum computing protocol comes from controlling the
distribution over the computing system's final state. Due to this, fidelity and
low operation cost are not in opposition, but go hand in hand, as Figs.
\ref{fig:tau_sweep} and \ref{fig:fid_work} demonstrate.

\subsection{Low Thermodynamic Cost}

\begin{figure}[t]
\centering
\includegraphics[width=\columnwidth]{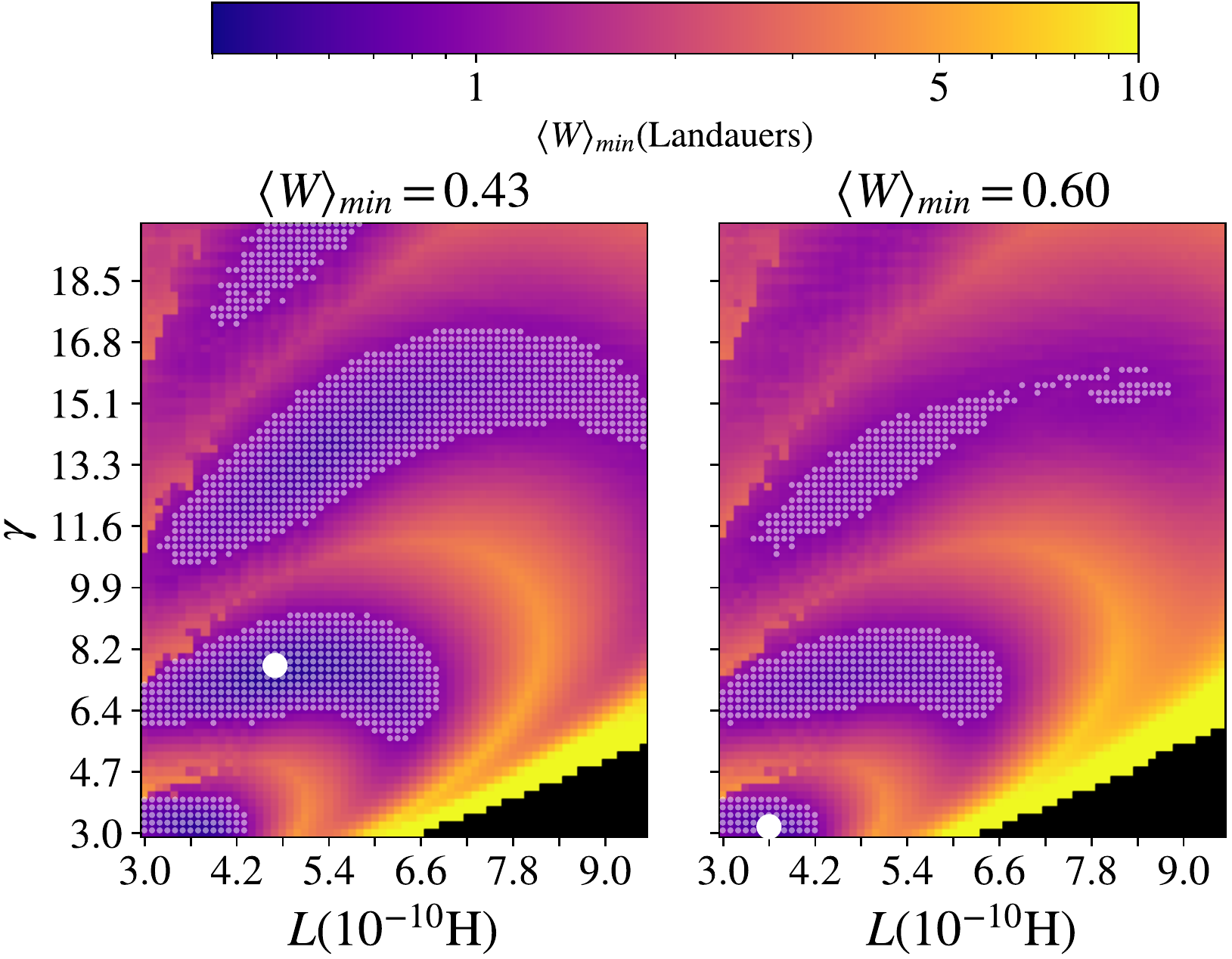}
\caption{Thermodynamic energy cost $\langle W \rangle_\text{min}$ for
	momentum-computing bit-swap over $5,120$ parameter combinations of $L$ and
	$\gamma$. (Left) Slightly asymmetric device with $I_- = \xIm$ gives the
	overall minimum $\langle W \rangle_\text{min} = 0.43$ Landauers (large
	solid white circle). (Right) Substantially asymmetric device with $I_- =
	\xImb$ gives the overall minimum $\langle W \rangle_\text{min} = 0.60$
	Landauers (large solid white circle). (Both) Small white circles indicate
	parameter values with protocols yielding $\langle W \rangle_\text{min} < 1$
	Landauer.  Black squares (lower right in each) represent parameter values
	where no successful swap was accomplished. Note that when the asymmetry is
	low, it can effectively be offset by the parameter $\px$, but for higher
	asymmetry, protocols that cost less than $1$ Landauer are less common.
	}
\label{fig:heatmaps}
\end{figure}

Conventional computing, based on transistor-network steady-state currents,
operates nowhere near the theoretical limit of efficiency for logical gates.
Even gates in Application Specific Integrated Circuits (ASICs) designed for
maximal efficiency operate on the scale of $10^4-10^6$ Landauers
\cite{chen2014, hamerly2019}. The physically-calibrated simulations described above
achieved average costs well below a Landauer for a wide range of parameter
values with an absolute minimum of $\langle W \rangle_\text{min} = 0.43$
Landauers, as shown in Figure \ref{fig:heatmaps} (left). For the less-ideal
asymmetric critical-current device (right panel), the cost increases to only
$\langle W \rangle_\text{min} = 0.60$ Landauers. And, the bulk of the
protocols we explored operated at $< 10$ Landauers. Altogether,
the momentum computing devices operated many orders of magnitude lower than the
status quo. Moreover, the wide basins reveal robustness in the
device's performance: an important feature for practical optimization and implementation.

\subsection{High Speed}

Paralleling accuracy, the now-conventional belief is that computational work
generally scales inversely with the computation time: $W \sim 1/\tau$
\cite{Boyd18a, zulkowski2015, aurell2012, reeb2014}. Again, this is not the
case for momentum computing, as Figs. \ref{fig:tau_sweep} and
\ref{fig:fid_work} demonstrate. Instead, there are optimal times $\tau^*$ that
give local work minima and around which the work cost increases. 

Optimal $\tau^*$s are upper bounded: the devices must operate \emph{faster}
than particular timescales---timescales determined by the substrate physics.
The bit swap's low work cost requires operating on a timescale faster than the
rates at which the system exchanges energy and information with the
environment. Thus, momentum computing protocols have a \emph{speed floor}
rather than a speed limit.

However, even assuming perfect thermal isolation there is a second bound on
$\tau^*$. The computation must terminate before the initially localized
ensemble---storing the memory---decoheres in position space due to dispersion.
For our JJ device this is the more restrictive timescale. Due to local
curvature differences in the potential, the initially compact state-space
regions corresponding to peaks of the storage potential's equilibrium
distribution begin to decohere after only one or two oscillations. Once they
have spread to cover both memory states, the stored information is lost. This
means it is most effective to limit the duration of the swap to just a
half-oscillation of the $\p$ coordinate. For our devices, this typically
corresponds to operating on timescales $< 15$ ns.

\section{Related Work}
\label{sec:litreview}

Reversible computing implementations of various operations have been proposed
many times over many decades. Perhaps the most famous is the Fredkin billiards
implementation \cite{Fred82a}. While ingenious, it suffers from inherent
dynamical instability (deterministic chaos) and cannot abide any interactions
with the environment. At the other end of the spectrum is a family of
superconducting adiabatic implementations \cite{likharev1982, likharev1996,
Take13a,Take13b,Take14a, soloviev2017, soloviev2018, schegolev2016}. These are
low cost in terms of dissipation and are stable, but they suffer from
fundamental speed limits due to the adiabaticity requirement: $\langle W
\rangle \propto 1/\tau$.

Other recent implementations \cite{osborn2019, frank2017, frank2019} of
reversible logic using JJs are more akin to the proposal at hand, in that they
require nearly-ballistic dynamics and attempt to recapture the energy used in a
swap at the final step. While these implementations are markedly different,
their motivation follows similar principles. Particularly, the framework for
asynchronous reversible computing proposed in  \cite{frank2017, frank2019}
might serve as a testbed for momentum computing elements.

Another distinguishing feature of the present design is that the phenomenon
supporting the computing is inherently linked to microscopic degrees of freedom
evolving in the device's phase space. This moves one closer to the ultimate
goal of using reversible nanoscale phenomena as the primitives for reversible
computing---a goal whose importance and difficulty were recognized by Ref.
\cite{morita2009}. Working directly with the underlying phase space also allows
incorporating the thermal environment. And, this facilitates characterizing the
effect of (inevitable) imperfect isolation from the environment.

It is worth noting the similarity between the optimal timescales $\tau^*$ and
the principal result in Ref. \cite{pidaparthi2021} in which a similar local
minima emerges when comparing thermodynamic dissipation to computation time.
These minima also come from certain matching conditions between the rate of
thermalization and the system's response time to its control device. Another
qualitatively similar result \cite{pankratov2004} found faster operation could
lead to reduced errors in overdamped JJs under periodic driving. These
similarities could point to a more general principle at play.

\section{Conclusion}
\label{sec:conclusion}

Our detailed, thermodynamically-calibrated simulation of microscopic trajectories
demonstrated that momentum computing can reliably (i) implement a bit swap at
sub-Landauer work costs at (ii) nanosecond timescales in (iii) a
well-characterized superconducting circuit. 

These simulations served two main purposes. The first highlights momentum computing's
advantages. The proposed framework uses the continuum of momentum states to serve as the
auxiliary system that allows a swap. In doing so, it eliminates the associated tradeoffs between
energetic, temporal, or accuracy costs that are commonly emphasized in
thermodynamic control analyses \cite{berut2012, Boyd18a, lahiri2016,
zulkowski2014}. Momentum computing protocols are holistic in that low
energy cost, high fidelity, and fast operation times all come from matching
parallel constraints rather than competing ones.

The second purpose points out key aspects of the proposed JJ circuit's
physics. The simulations reveal several guiding principles---those that
contribute most to decreasing work costs for the proposed protocols. The
system is so underdamped that thermal agitation is not the primary cause of inefficiency. The two main contributors are (i) the appearance of
dispersive behavior in the dynamics of an initially-coherent region of state
space and (ii) asymmetries inherent to the device that arise from differing
critical currents in the component superconducting JJ elements. Notably, if the
elements are very close to each other in $I_c$, then symmetry can be
effectively restored by setting the control parameter $\px$ to counteract the
difference. However, the more asymmetry, the harder it is to find ultra
low-cost protocols; cf. Fig. \ref{fig:heatmaps} left and right panels. Note, too, that initial-state dispersion can be ameliorated by using a
$\VC$ that is as harmonic (quadratic) as possible. However, this typically
requires lower inductance $L$, possibly complicating circuit fabrication.
Additionally, the potential-well separation parameter $\beta$'s linear
dependence on $L$ hinders the system's ability to create two distinct states
during information storage. Though these tradeoffs are complicated, our simulations suggest that dispersion can be controlled, yielding swap
protocols with even lower work costs.

Since the protocol search space is quite high-dimensional and contains many local-minima,
we offer no proof that the protocols found give the global work minimum. Very likely, the thermodynamic costs and operation speed of our proposed JJ
momentum computing device can be substantially improved using more
sophisticated parameter optimization and alternative materials.  Even with the
work cost as it stands, though, sub-Landauer operation represents a radical
change from transistor-based architectures. One calibration for this is given
in the recent stochastic thermodynamic analysis of a NOT gate composed of
single-electron-state transistors \cite{gao2021} that found work costs
$10^4$ times larger.

Note, too, that running at low temperatures requires significant off-board
cooling costs, as required in superconducting quantum computing. Our current
flux qubit implementation requires operating at liquid He temperatures
\cite{saira2020, Wims19a}. However, there are also JJs that operate at $N_2$
temperatures, promising system cooling costs that would be $2$ to $3$ orders of
magnitude lower \cite{Yurg00a,Long12a,cybart2015,Revi21a}.

Additionally, the physics necessary to build a momentum computing
swap---underdamped behavior and controllable multiwell dynamics---is far from
unique to superconducting circuits. As an example, nanoelectromechanical
systems (NEMS) are another well-known technology that is scalable with modern
microfabrication techniques. NEMS provide the needed nonlinearity for
multiple-well potentials, are extremely energy efficient, and have high Q
factors even while operating at room temperature \cite{Lifs08a,Math14a,Ryu21a}.
Momentum computing implemented with NEMS rather than superconductors completely
obviates the cooling infrastructure and so may be better suited for large-scale
implementations.

That said, the JJ implementation at low temperatures augmented with appropriate
calorimetry will provide a key experimental platform for careful, controlled,
and detailed study of the physical limits of the thermodynamic costs of
information processing. Thus, these devices are necessary to fully understand
the physics of thermodynamic efficiency. And so, beyond technology impacts, the
proposed device and protocols provide a fascinating experimental opportunity to
measure energy flows that fluctuate at GHz timescales and at energy scales
below thermal fluctuations. Success in these will open the way to theoretical
investigations of the fundamental physics of information storage and
manipulation, time symmetries, and fluctuation theorems \cite{riechers2020,
boyd2021}.

\section*{Acknowledgments}
We thank Alec Boyd, Warren Fon, Scott Habermehl, Jukka Pekola, Paul Riechers,
Michael Roukes, Olli-Pentti Saira, and Gregory Wimsatt for helpful discussions.
The authors thank the Telluride Science Research Center for hospitality during
visits and the participants of the Information Engines Workshops there. JPC
acknowledges the kind hospitality of the Santa Fe Institute, Institute for
Advanced Study at the University of Amsterdam, and California Institute of
Technology. This material is based upon work supported by, or in part by, FQXi
Grant number FQXi-RFP-IPW-1902 and U.S. Army Research Laboratory and the U.S.
Army Research Office under grants W911NF-21-1-0048 and W911NF-18-1-0028.

\appendix

\begin{table*}[ht]
\begin{tabular}{|l|c|c|}
\hline
Environment & Temperature $T$ & Thermodynamic Energy \\
  $\text{}$ & Kelvin($K$)  & Joules($J$) \\
\hline
Microprocessor       & 373    & $5.2 \times 10^{-21}$\\
Room Temp            & 293    & $4.0 \times 10^{-21}$ \\
Liquid $N_2$         &  77    & $1.1 \times 10^{-21}$ \\
Liquid $He$          &  4.2   & $5.7 \times 10^{-23}$ \\
$1$ K                &  1.0   & $1.4 \times 10^{-23}$ \\
$1$ mK            &  0.001 & $1.4 \times 10^{-26}$ \\
\hline
\end{tabular}
\caption{Thermodynamic energy in environments at various temperatures.
	}
\label{tab:ThermoEnergy}
\end{table*}

\newcolumntype{d}[1]{D{.}{.}{#1}}

\begin{table*}[ht]
\begin{tabular}{|l|c|c|c|}
\hline

 Operation& Landauers ($L$) & Environment $T$ & Energy \\
   $\text{}$  & $\text{}$ & Kelvin ($K$) & Joules ($J$)\\
\hline
CMOS gate \cite{gao2021}
                 & $7000$    & 293    & $1.9 \times 10^{-17}$ \\
CMOS gate \cite{freitas2021}
                 & $3000$    & 293    & $8.4 \times 10^{-18}$ \\
CMOS bound \cite{frank2005,ITRS20a}
                 & $100$     & 293    & $2.8 \times 10^{-19}$ \\
Bit Erase (Ideal) \cite{Land61a}
                 & $1$      & 293    &   $2.8 \times 10^{-21}$ \\
Bit Erase (Ideal) \cite{Land61a}
                 & $1$    &  1     &   $9.6 \times 10^{-24}$ \\
Bit Swap (JJ)    & $0.43$    &  1   &  $4.1 \times 10^{-24}$ \\
Bit Swap (Ideal)    & $0$    &  293   &  $0$ \\
Bit Swap (Ideal)    & $0$    &  1   &  $0$ \\
\hline
\end{tabular}
\caption{Landauers and work energies (Joules) for various information processing
	operations in environments and at temperatures where thermodynamic
	computers may operate.
	}
\label{tab:Landauers}
\end{table*}

\section{The Landauer: A Standard Candle for Thermodynamic Computation}
\label{sm:Landauer}

A long and checkered history underlies the physics of information and energy,
arguably originating in the paradox of Maxwell's Demon \cite{Leff02a}. Most
recently, though, the paradigm of \emph{thermodynamic computing} emerged to
frame probing their limits \cite{Cont19a}. In this setting, Landauer's
Principle says that $\kB T \ln 2$ energy units must be expended to erase a
single bit of information. Beyond erasure, though, his Principle also stands as
a challenge---Can conventional computing paradigms operate at sub-Landauer
scales? It seems not. Landauer's theory and follow-on results
\cite{Boyd15a,parrondo2015,Wims20a} and recent experiments \cite{berut2012,Jun14a} verified the lower bound.

To apply more broadly, \emph{Landauer's Principle} generalizes to $W \geq \kB T
\Delta H$, where $\Delta H$ is the change in Shannon entropy
between a computational system's initial and final information-bearing states
\cite{landauer1961,parrondo2015,deffner2013}. Despite the Principle's
generalization beyond bit erasure, the Landauer scale remains a familiar
reference point for the energy costs of binary operations; its familiar use
coming at the expense of ignoring specifics of any given logical operation
\cite{gammaitoni2011}. 

An efficient bit-swap operation, for example, has zero generalized Landauer
cost, as it is logically reversible. However, since many thermodynamic
computing architectures do not have access to dynamics that can accomplish
reversible computing efficiently, the Landauer scale provides a common
reference to compare gate performance across physical substrates and design
paradigms. It also facilitates comparing across substrates that operate at
different temperatures. Table \ref{tab:ThermoEnergy} lists thermodynamic
energies for a range of physical environments. Table \ref{tab:Landauers} gives
Landauer work energies for various information processing operations in
environments and at temperatures where thermodynamic computers operate.

\section{Limits of Stochastic Thermodynamics for Information Processing}
\label{sm:LimitStochThermo}

Stochastic thermodynamics \cite{esposito2012, seif2019} has been the
predominant framework for analyzing the thermodynamic costs of stochastic
mappings. It assumes the memory state $m$ obeys stochastic Markovian dynamics:
continuous-time Markov chains (CTMCs), where the state distribution
$\vec{p}(t)$ changes continuously as a function of itself: $\dot{\vec{p}}(t)=
f(\vec{p},t)$. The resulting dynamics are necessarily represented by a master
equation over the memory-state distribution $\dot{\vec{p}}(t)=\mathbf{A}(t)
\vec{p}(t)$ \cite{owen2019}. This framework is powerful, yielding great insight
into physical processes when its assumptions are met.

The framework, however, does not apply to momentum computing. To appreciate
why, consider justifying Markovian dynamics over memory states. Assume a
microscopic physical system $\mathcal{S}$ that serves as a computational
substrate. While allowing the universe to be deterministic, $\mathcal{S}$ can
exhibit stochastic dynamics since it represents only a portion of the
partially-observed universe. The very typical assumption that $\mathcal{S}$'s
local environment acts as a large weakly-coupled heat bath with quickly
relaxing degrees of freedom yields dynamics on $\mathcal{S}$ that are also
Markovian and, therefore, can be represented by CTMCs.

However, computationally-useful memory states are not the CTMC-obeying
microstates of $\mathcal{S}$, but a set $\mathcal{M}$ of mesostates that
represent coarse-graining over $\mathcal{S}$. It is possible, depending on the
variables or timescales of interest, that this coarse-graining ignores only
rapidly-relaxing subsystems of $\mathcal{S}$. Then $\mathcal{M}$ inherits the
Markov property that governs the microstates \cite{esposito2012}. This
strategy---coarse graining over physical degrees of freedom irrelevant to the
dynamics---is analogous to establishing $\mathcal{S}$ as a stochastic,
Markovian subsystem of the universe. A straightforward example of this case is
when $\mathcal{M}$ consists of positional degrees of freedom and $\mathcal{S}$
evolves by overdamped Langevin dynamics.

When implementing momentum computing, however, the coarse-graining yielding
$\mathcal{M}$ is applied over \emph{hidden microstates} that contain
dynamically relevant information not determined from $\mathcal{M}$'s
instantaneous realizations. As a consequence, the dynamic over the
coarse-grained states is not Markovian. CTMCs cannot be used. A straightforward
example of this arises when $\mathcal{M}$ consists of positional degrees of
freedom and $\mathcal{S}$ evolves by underdamped Langevin dynamics. On the
downside, a general analytical treatment of such partially-observed systems
(continuous-time \emph{hidden} Markov chains) is highly nontrivial
\cite{koyu19, seif2019, maes17, strasberg2019}. On the upside, the possibility
of hidden states allows for substantially more general forms of computation. As
the results here showed, the benefits of this expanded space are quite
substantial.

Note that the bit swap computation is, in general, problematic to implement
using CTMCs since input-output mappings whose determinants are negative are
disallowed when memory-state dynamics are restricted to obey CTMCs. Formally,
auxiliary systems can be added to the set of memory states. Done correctly this
again permits using CTMCs in the augmented state space to accomplish the
computation \cite{owen2019}.

However, physically-embedded computations do not generally allow the required
perfect control over the system Hamiltonian. Indeed, one need look no further
than the present work to see how nontrivial it is to implement an operation as
simple as a harmonic oscillation in a physically-realistic device.

Moreover, adding auxiliary subsystems increases state-space dimension and
complicates control apparatus and control protocols. Due to the increased
complication, in many settings, adding auxiliary dimensions is simply not
physically possible. On top of this, the timescale of these augmented
computations must be longer than the equilibration time of the auxiliary
systems and thermal environment. In this way, adding auxiliary systems imposes
additional speed limits to computations. In short, adding auxiliary subsystems
 addresses the shortfalls of CTMCs, but does not sidestep their fundamental limitations.
 
We illustrate this by considering an efficient bit swap implemented via a Markovian embedding. First, it augments the system with an unoccupied auxiliary state $A$
to serve as a transient memory. It then quasistatically translates memory state
$0$ to $A$, while memory state $1$ is translated to $0$. Finally, it
quasistatically translates $A$ to $1$.

Quasistatic processes cost arbitrarily little work, but they take
arbitrarily-long times. To compute faster ($\tau \to 0$), the work cost will
diverge as $1/\tau$ \cite{Boyd18a, zulkowski2015, aurell2012, reeb2014}. Increasing fidelity requires raising the
scale of the barrier separating the states. Doing so, though, increases the
energetic cost at a given computational speed; maintaining the same
work cost, then, requires slowing the operation. In short, the trade-offs in Markovian
embedding complicate design and, more to the point, reduce performance.

\section{Flux Qubit Dimensionless Equations of Motion}
\label{sm:DimensionlessEoM}

\def \df#1{\widehat{#1}}
\def \dl#1{#1}

In terms of the dimensional degrees of freedom, the flux qubit equations of
motion are:
\begin{align}
\ddot{\df\varphi} & = -\frac{2}{RC} \dot{\df\varphi} -\frac{1}{C} \partial_{\df\varphi} U(\df\varphi, \df\varphi_{dc})
    \\
\ddot{\df\varphi}_{dc} & = -\frac{2}{RC} \dot{\df\varphi}_{dc} - \frac{4}{C}\partial_{\df\varphi_{dc}} U(\df\varphi, \df\varphi_{dc})
  ~,
\end{align}
where the dimensional $\df \varphi$s are related to the main text's
dimensionless fluxes and phases by the magnetic flux quantum $2\pi / \Phi_0$.
With the addition of thermal noise, the Langevin equation is:
\begin{align}
dv_i = -\frac{\nu_i}{m_i} v_i dt - \frac{1}{m_i} \partial_{x_i} U(x) dt + \frac{1}{m_i} r(t)\sqrt{2\nu_i \kappa dt}
  ~,
\end{align}
where $\kappa\equiv k_B T$. Matching these variables to the equations of motion
yields:
\begin{align}
x &=  \left(\df\varphi, \df\varphi_{dc}\right)  \\
v &= \left(\dot{\df\varphi}, \dot{\df\varphi}_{dc}\right) \\
m &= \left( C, \frac{C}{4}\right), ~\text{and} \\
\nu &= \left(\frac{2}{R}, \frac{1}{2R}\right)
  ~,
\end{align}
where subscript $i$ has been dropped in favor of a vector representation.

The task is to write each physical quantity $z$ in terms of a dimensional
constant and dimensionless variable by defining scaling factors according to
the following prescription: $ z \equiv z' z_c$, where $z_c$ is a dimensionful
constant.

Setting $m_c = C$ and $\nu_c = 1 / R$ are obvious choices. Additionally,
since the potential factors into $U = U_0 \times U'(\frac{2\pi}{\Phi_0} \cdot
x)$, a good choice for positional scaling is $x_c = \Phi_0 / 2\pi$.

It is advantageous to write nondimensional kinetic energies as $\frac{1}{2} m'
v'^2$ without additional scaling factors. This means setting the energy scaling
as:
\begin{align}
E_c  = m_c \frac{ x^2_c}{t^2_c}
  ~.
\end{align}

This does not uniquely determine the energetic scale, since $t_c$ is still
free. The two obvious choices are to scale to the temperature---$KE'=1$
corresponds to $k_B T$ units of dimensional energy---or to the potential energy
scale---$KE'=1$ corresponds to $U_0$ units of dimensional energy. Choosing
the latter yields:
\begin{align}
E_c & = U_0 = m_c \frac{ x^2_c}{t^2_c} \text{~and}\\
\frac{x_c^2}{L} & = m_c \frac{ x^2_c}{t^2_c}
  ~.
\end{align}
Evidently, the timescale is $t_c = \sqrt{LC} $, which is a workable timescale
for our purposes given that the dynamics of interest happen on the scale of
$\tau \approx \omega_{LC}$. Setting the timescale to the potential energy
rather than the thermal energy may well become common practice in simulating
momentum computation, since protocols must be timed precisely with respect to
the dynamics of the potential energy surface.

The Langevin equation, in terms of the nondimensional quantities defined above,
becomes:
\begin{align}
dv' \frac{x_c}{t_c} &= -\frac{\nu' \nu_c}{m' m_c} v' x_c  dt'
 - \frac{1}{m' m_c} \left(\frac{U_0}{x_c} \partial_{x'} U'(x')\right)
   t_c dt' \\
  & \quad + \frac{1}{m'm_c} r(t)\sqrt{2\nu'\nu_c E_c \kappa' t_c dt'} \nonumber
  ~.
\end{align}
Simplifying algebra then yields:
\begin{align}
dv' &= -\frac{\sqrt{LC} }{ RC} \frac{\nu'}{m'} v' dt'
  - \frac{1 }{ m'}  \partial_{x'} U'(x') dt' \\
  & \quad + \left(  \frac{L}{R^2 C} \right)^{1/4} \frac{\sqrt{\nu'\kappa'}}{m'} r(t) \sqrt {2 dt'} \nonumber
  ~.
\end{align}
Finally, we define $\lambda$, $\theta$, and $\eta$ as nondimensional parameters
that serve as our dimensionless Langevin coefficients. This yields the Langevin
equation for the simulations detailed in Appendix \ref{sm:SearchMinWork}:
\begin{align}
dv' = -\lambda v' dt' - \theta \partial_{x'} U' + \eta r(t) \sqrt{2dt'}
  ~,
\end{align}
with:
\begin{align}
\lambda & = \frac{\sqrt{LC} }{ RC} \frac{\nu'}{m'}, \\
\theta & = \frac{1 }{ m'}, ~\text{and} \\
\eta & = \sqrt\frac{\lambda \kappa'}{m'}
	~,
\end{align}
where:
\begin{align}
x' &= (\dl\varphi, \dl\varphi_{dc}), \\
v' &= \frac{d}{dt'} x', \\ 
\nu' &= (2, 1/2), \\
m' &= ( 1, 1/4), ~\text{and} \\
\kappa' &= \frac{k_B T}{U_0}
    ~.
\end{align}

\section{Effective Potential and Simulation Details}
\label{sm:PotentialSimplifications}

We consider two cases: critical-current symmetric and asymmetric JJ pairs.

\subsection{Symmetric Approximation}

We can obtain reasonable estimates for good $\pxdc$ values by assuming a
perfectly symmetric device $\delta\beta=0$. Furthermore, we also set $\px =0$
for all cases. This allows two symmetric wells on either side of $\p =0$. In
practice, since $\delta \beta \neq 0$ in a real device, $\px$ would be
calibrated to compensate for the asymmetry; see Sec. \ref{sec:dbneq0}.

In the symmetric case, the potential splits into two components---periodic and
quadratic:
\begin{align}
\beta \cos \p \cos \frac{\pdc}{2} +  \frac{1}{2} \p^2 + \frac{\gamma}{2} (\pdc-\pxdc)^2
  ~.
\end{align}
The periodic term allows for multiple minima, while the quadratic terms force
the dynamical variables to stay close to their respective parameters. This
localization means we focus only on the the area near $\p = \px$ and $\pdc =
\pxdc$.

To employ the potential most flexibly, we must characterize the relevant fixed
points that occur in this region. Following Refs. \cite{saira2020,
han1992experiment}, we choose to search in the domain $-\pi < \p < \pi$ and
$-2\pi < \pdc < 0$. Fixed points occur when all components of the gradient
vanish:
\begin{align}
\label{eq:grad1}
 \partial_\p U' &= -\beta \sin\p \cos \frac{\pdc}{2} + \p = 0 \\
 \label{eq:grad2}
 \partial_{{\pdc}} U' &= -\frac{\beta}{2} \sin \frac{\pdc}{2} \cos \p + \gamma ( \pdc - \pxdc) =0
\end{align}

The first condition is met whenever $\p=0$ and, also, when $\frac{\p}{\beta
\sin \p} = \cos \frac{1}{2} \pdc$. Consider the case where $\p=0$---the
``central'' fixed point. To find the $\pdc$ location of the fixed point
$\p^0_{dc}$, we look to the gradient's second term. This yields the condition: \begin{align}
\label{eq:p0}
\pdc^0 - \frac{\beta}{2\gamma} \sin \frac{\pdc^0}{2} & = \pxdc
\\ 
\nonumber
F^0(\pdc=\pdc^0, \beta, \gamma) & = \pxdc
  ~.
\end{align}
The central fixed point occurs close to the parameter $\pxdc$, but is offset by
a value $\leq \beta / 2\gamma$.

The equation above can be solved numerically with ease to find the location of
the central fixed point. To classify the fixed point, we look at the Hessian.
While the general expression for the eigenvalues is rather verbose, the case
where $\p=0$ simplifies to:
\begin{align}
\lambda_1 & = -\beta\cos\frac{\pdc^0}{2} +1 \\
  \lambda_2 & = \gamma -\frac{\beta}{4} \cos\frac{\pdc^0}{2} 
  ~.
\end{align}

\begin{figure}[t]
\centering
\includegraphics[width=\columnwidth]{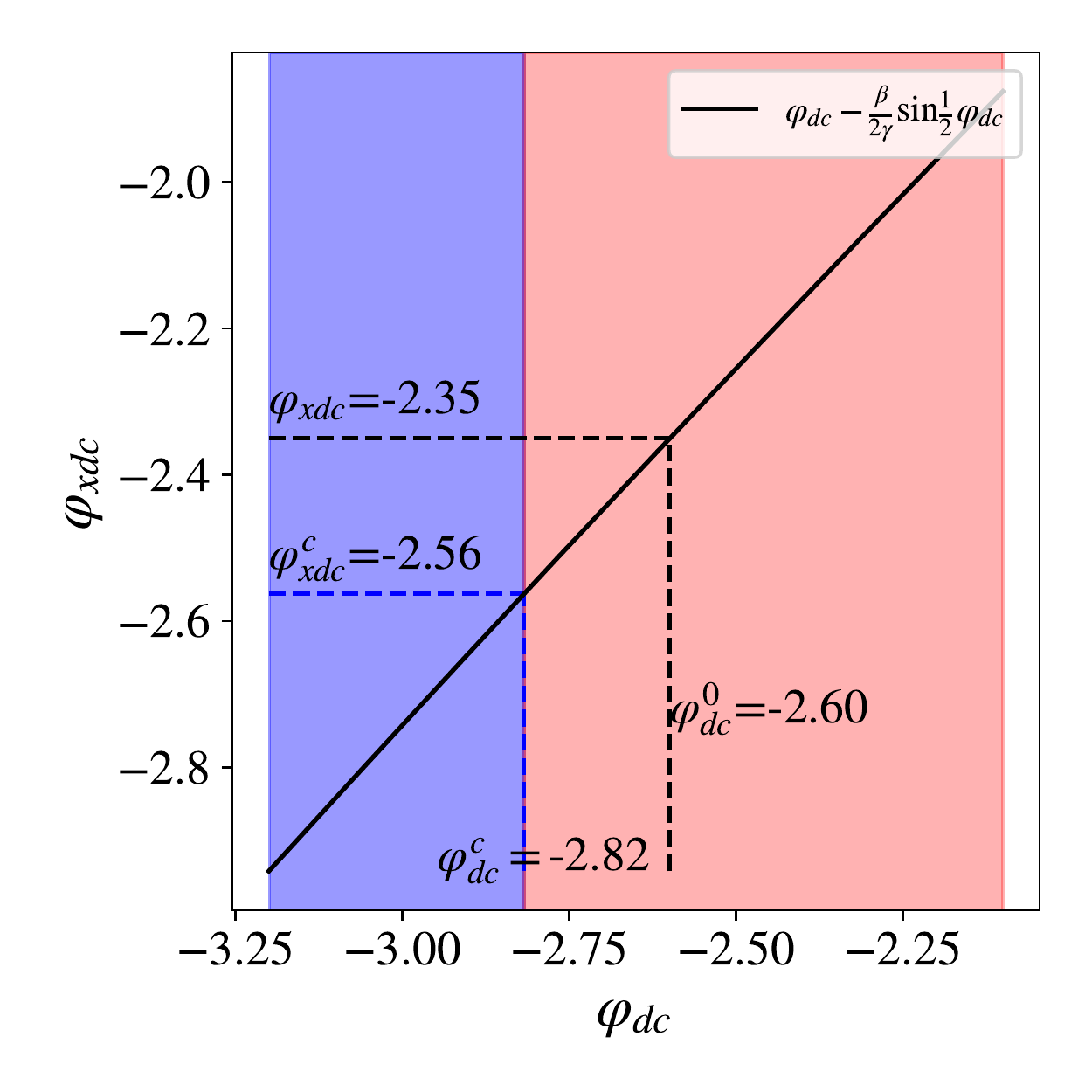}
\caption{Fixed point at $\p=0$ in an ideal device with $\beta = 6.2$ and
	$\gamma = 12.0$: Red (Blue) background indicates regions where the fixed
	point is a saddle point (local minimum). For example, if $\pxdc = -2.35$,
	the central fixed point is a saddle point at $\pdc = -2.6$. To find a
	stable fixed point at $\p=0$, a control parameter less than $\pxdc^c$ is
	necessary, which falls at $-2.56$ in the example above.
	}
\label{fig:p0}
\end{figure}

$\lambda_2 >0$ as long as $\gamma > \beta / 4$. And, since we assume $\gamma >
\beta$, this condition is always met. Thus, this fixed point is either a saddle
point or a minimum based on whether $\pdc^0$ is greater or less than $ \pdc^c
\equiv -2 \cos^{-1}\frac{1}{\beta}$, respectively. (We only use the negative
branch of $\cos^{-1}$ due to the domain of $\pdc$.) See Fig. \ref{fig:p0} for
an example of the behavior of the central fixed point for typical parameters.

We can also find an expression for $\pxdc^c(\beta,\gamma) \equiv F^0(\pdc =
\pdc^c)$, the critical value of the control parameter at which the central
fixed point transitions between a saddle point and a minimum:
\begin{align}
\nonumber
\pxdc^c(\beta,\gamma) &= \pdc^c - \frac{\beta}{2\gamma} \sin \frac{\pdc^c}{2}
\\
& = -2\cos^{-1} \frac{1}{\beta} + \frac{\beta}{2\gamma} \sqrt {1-\frac{1}{\beta^2}}
  ~.
\label{eq:phi^c_xdc}
\end{align} 

Naively, the best strategy to form a low cost protocol is to take values of
$\pxdc$ just above and below $\pxdc^c$. However, there are several factors that
introduce complications. For one, the energy scale separating the two wells
when $\pxdc \approx \pxdc^c$ is very small and it will typically be overwhelmed
by thermal energy at the temperatures of interest ($400-1400$ mK). A second is
that the approximation of $\delta \beta =0$ actually has a most pernicious
effect near $\pxdc^x$. (This is discussed in Sec. \ref{sec:dbneq0}.)

Finally, we have yet to consider the other fixed points at $\p \neq 0$. Doing
so reveals that sometimes $\pxdc^c$ corresponds to a subcritical pitchfork
bifurcation---yielding a potential with a third (undesirable) minimum rather
than a single one.

When $\p \neq 0$ we can rewrite Eqs. (\ref{eq:grad1}) and (\ref{eq:grad2}):
\begin{align}
 \frac{\p}{\beta \sin \p} & = \cos \frac{1}{2} \pdc \\
 \frac{\beta}{4\gamma} \sin \frac{\pdc}{2} \cos \p - \frac{1}{2} \pxdc & = \frac{1}{2} \pdc
  ~.
\end{align}

\begin{figure}[t]
\centering
\includegraphics[width=\columnwidth]{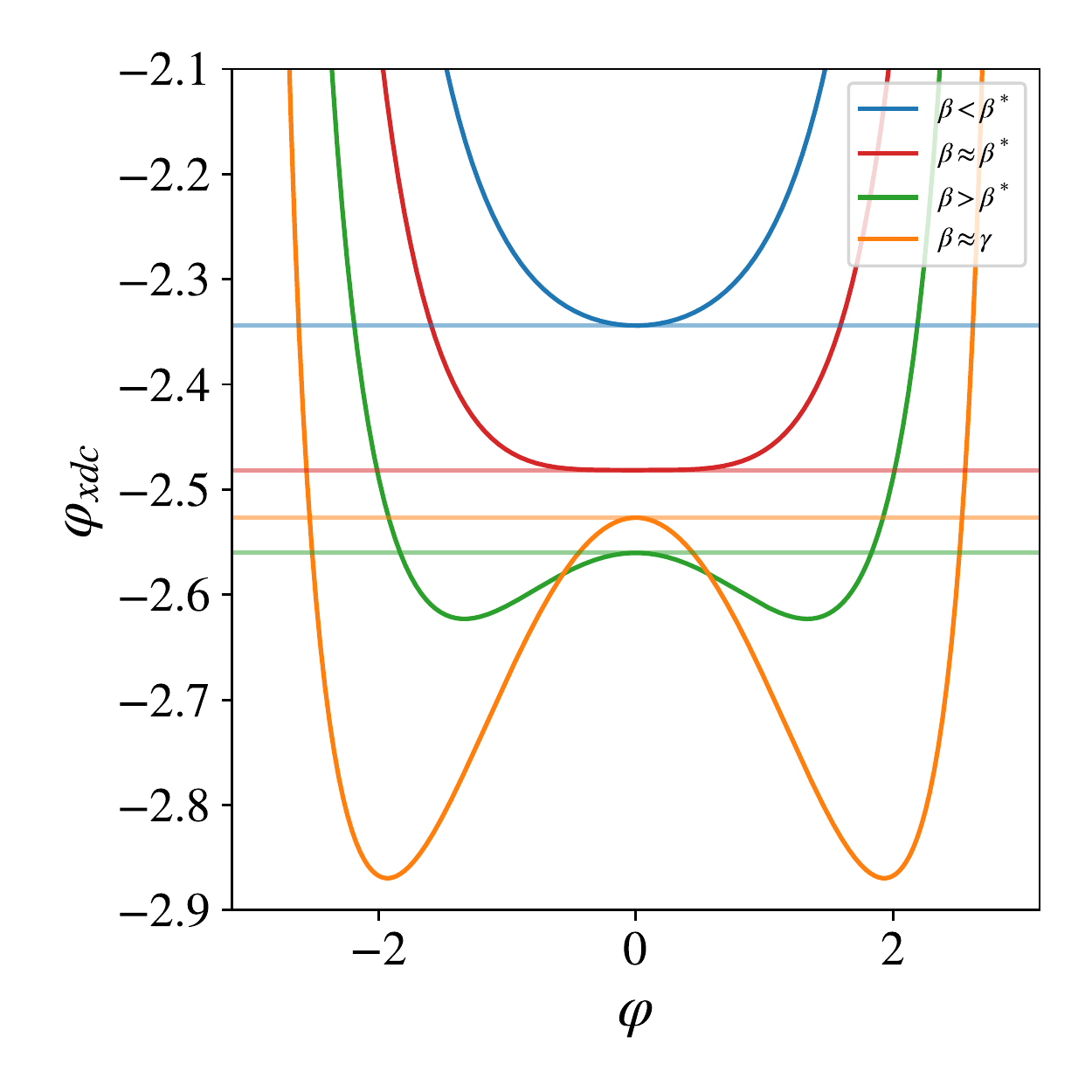}
\caption{$\p\neq0$ fixed points appear when the value of the function plotted
	equals the external $\pxdc$ parameter. Note that for some $\beta$ and
	$\gamma$ combinations, there is a qualitatively different behavior.
	Especially for larger $\beta$, there is a coexistence region of three
	potential minima. For the $\beta\approx\gamma$ example, one would want to
	set $\Delta C > 0.5$ to make sure $\VC$ falls well outside of the three
	minima range. Horizontal lines show the values of $\p^c_{xdc}$. (See Appendix
	\ref{sm:SearchMinWork}.)
	}
\label{fig:pneq0}
\end{figure}

The potential is symmetric, so these fixed points come in pairs $\p^{\pm}$.
Substituting $\pdc/2 = -\cos^{-1} (\p^{\pm} / \beta \sin \p^{\pm})$ into the
second equation yields the following for $\p^{\pm}$:
\begin{align}
\pxdc & = \frac{\beta}{2\gamma}
  \sqrt{ 1- \left(\frac{\p^{\pm}}{\beta \sin \p^{\pm}}\right)^2} \cos \p^{\pm}
  -  2\cos^{-1} \frac{\p^\pm}{\beta \sin\p^\pm} \\
 \pxdc & = F^{\pm} (\p=\p^{\pm}, \beta,\gamma)
  ~.
\end{align}
Note that the sign changes due to the domain restriction of $\pdc$. Figure
\ref{fig:pneq0} shows how these fixed points behave as $\beta$, $\gamma$, and
$\pxdc$ change. The value of $\pxdc$ tangent to the curve when $\p=0$
corresponds to the critical control parameter value $ \pxdc^c$, which can be
seen by verifying $\lim_{\p\to0} F^{\pm}(\p) = \pxdc^c$.

As a last note, different values of $\beta$ and $\gamma$ have qualitatively
different fixed point profiles depending on whether the central fixed point
undergoes a supercritical or subcritical pitchfork bifurcation when $\pxdc =
\pxdc^c$. The critical value $\beta^*$ where the bifurcation of the central
fixed point transitions between being supercritical and subcritical is given by:
\begin{align}
\lim_{\p\to0}\partial^2_\p F^{\pm}(\p, \beta^*, \gamma) = 0
  ~.
\end{align}
Once again, the full derivative is quite verbose. However, taking the limit $\p
\to 0$ gives:
\begin{align}
\frac{\sqrt{\beta^{*^2}-1}}{6\beta^{*^2}} \left( -3\beta^{*^2} + 4\gamma +
2\right) & = 0 \\
\beta^{*} & = \sqrt\frac{4\gamma+2}{3}
  ~.
\end{align}

\begin{figure}[t]
\centering
\includegraphics[width=\columnwidth]{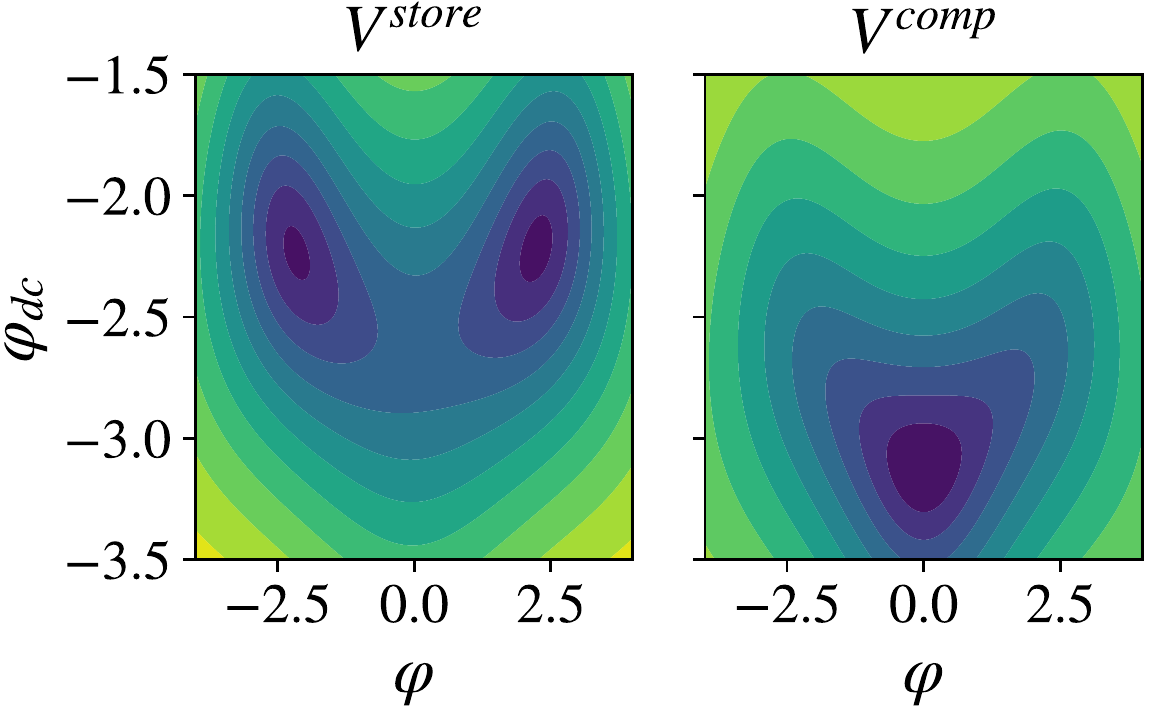}
\caption{(Left) $V^{\text{store}}$, the bistable storage potential.
	(Right) $V^{\text{comp}}$, the ``banana-harmonic'' potential. These potential energy profiles serve as qualitative pictures to represent prototypical computational and storage potentials, and do not represent any particularly favorable parameter set. 
	}
\label{fig:potentials}
\end{figure}

Interestingly, when $\beta>\beta^*$, there is always a parameter space region
with three distinct minima. This might be useful, in fact, for single-bit
computations that require more states. For bit swap, though, the goal is for
the system to jump between a $\VS$ with $2$ minima and a $\VC$ with a single
minimum (see Figure \ref{fig:potentials}). And so, care must be taken to avoid the three-minima regions when
$\beta > \beta^*$.

\subsection{$\delta\beta\neq0$}
\label{sec:dbneq0}

The device just considered is ideal. In reality $\delta\beta\neq0$, and exact
analytic work is much less fruitful. Introducing the asymmetric terms augments
the potential:
\begin{align}
U_{\text{asym}} (\p,\px,\delta\beta,\pdc)= \frac{1}{2} \px^2 - \p \px - \delta\beta \sin\p\cos \frac{\pdc}{2}
  ~.
\end{align}
In short, one must vary $\px$ to offset the effect of $\delta\beta$, provided a
symmetric potential is preferred.

There are two obvious strategies to minimize the effects of asymmetry. Either a
strategy that minimizes the effect of $U_{asym}$ at the central fixed
point---the ``min of mid'' strategy---or at the fixed points at $\p^\pm$---the
``min of max'' strategy. It stands to reason that one uses the former to set
$\px$ for $\VC$ and the latter for $\VS$.

The ``min of mid'' strategy is easy to implement. Simply set the derivative of
$\partial_{\p} U_{asym} |_{\p=0} =0$, with the intent of having the
asymmetrical part of the potential be as flat as possible near $\p=0$. Simple
algebra yields: $\px = -\delta\beta \sin \pdc/2$.

The ``min of max'' strategy requires numerical solution. First, note that the
maximum value of $U_{asym}$ occurs when $\p= \p_{max} = \arccos
(\frac{\px}{\delta\beta \sin .5 \pdc})$. Then, use a symbolic solver (e.g.,
SymPy's \emph{nsolve} function) to find the value of $\px$ that minimizes
$U_{asym}(\p_{max}, \px, \delta\beta, \pdc)$.

\begin{figure}[t]
\centering
\includegraphics[width=\columnwidth]{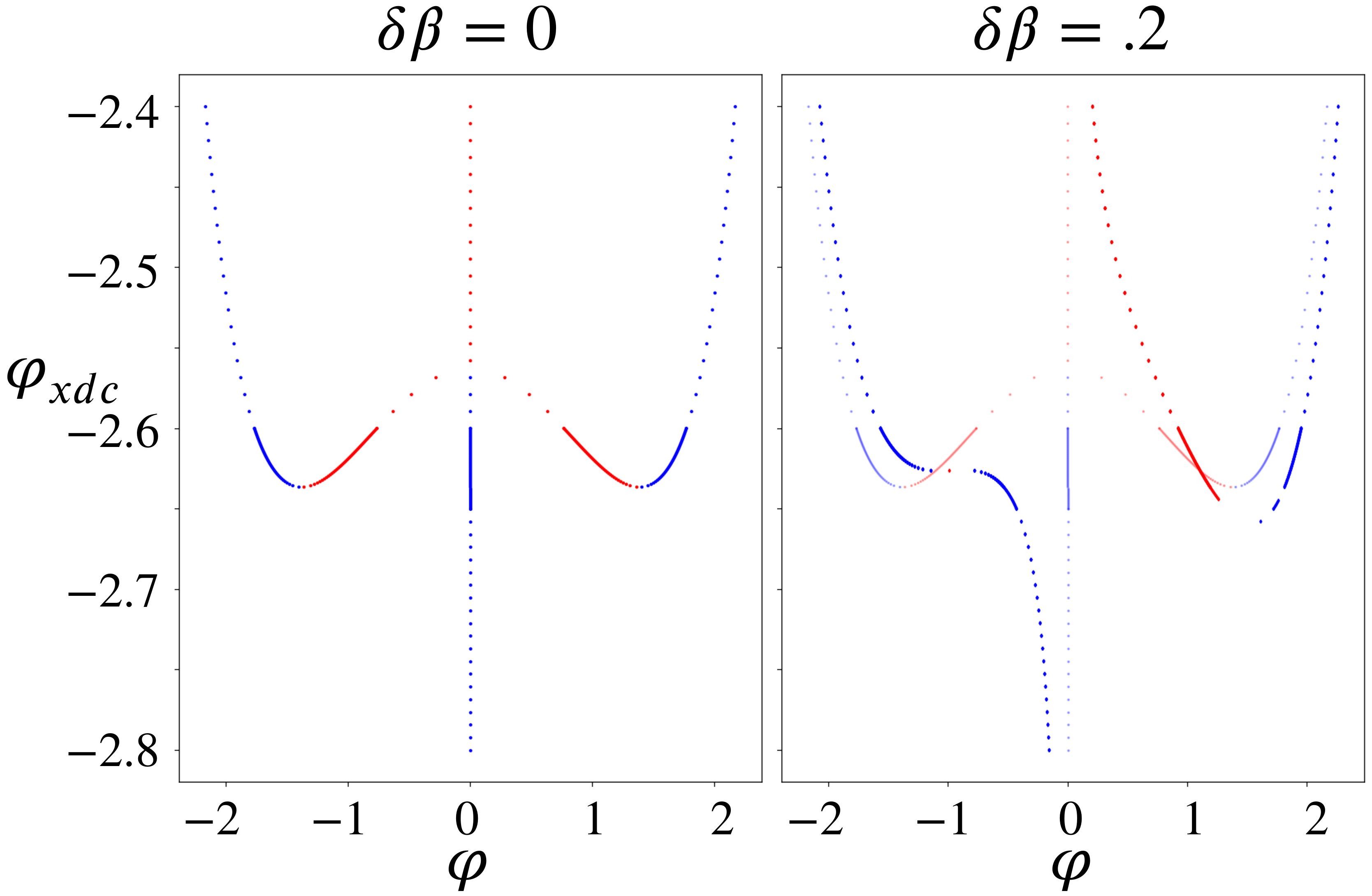}
\caption{Fixed point bifurcation diagram for the (left) idealized
	$\delta\beta=0$ device and (right) a device with $\delta\beta=0.2$. Blue
	indicates stable minima and red saddle points. On the right plot, the
	$\delta\beta=0$ fixed points are plotted as well, with low opacity to help
	see the difference. The naive ``minimum of maximum'' strategy has been used
	to minimize the effect of $U_{asym}$. And, we can see that the symmetric
	approximation works fairly well as long as $|\pxdc-\pxdc^c| > .2$. It is
	likely that more evolved solution strategies will improve results.
	}
\label{fig:ideal_vs_exact}
\end{figure}

Figure \ref{fig:ideal_vs_exact} shows that the effect of $\delta\beta \neq 0$
is, unsurprisingly, the most noticeable near the bifurcation of the central
fixed point. For the bit swap, as described in Sec. \ref{sec:exact_protocol},
we need only two different profiles for the potential: one in which we have two
symmetric wells and one in which we have a single well placed midway between
them. Thus, we must keep the $\pxdc$ parameter sufficiently far away from
$\pxdc^c$. The strategy employed in the simulations described below always
involves setting a minimum distance that $\pxdc$ must be from $\pxdc^c$, in
order to avoid falling into the pitfalls described here.

\section{Searching for Minimal-Work Bit Swaps}
\label{sm:SearchMinWork}

The following lays out the computational strategy to find low work-cost
implementations.

We are most interested in the effect of parameters that are the most removed
from fabrication, so all simulations assume JJ elements with $I_+$, $R$, and
$C$ set to $\xIp$, $\xR$, and $\xC$, respectively. To explore how asymmetry
affects work cost, we simulated protocols with a nearly-symmetric device with
$I_- = \xIm$, a moderately-symmetric device with $I_- = \xImb$, and an
asymmetric device with $I_- = \xImbt$. Additionally, $\kB T$ is always scaled
to $U_0$, so that $\kappa' \equiv \kB T / U_0 = 0.05$.

Given devices with the parameters above, what values of the remaining
parameters yield protocols with minimum work cost? This involves a twofold
procedure. First, create the circuit architecture by setting $L$ and $\gamma$ by
hand; thus, fully specifying the device. Second, determine the ideal protocols
for that combination of device parameters through simulation.

$L$'s order of magnitude was chosen from previous results \cite{barone1982,
han1992theory, han1992experiment, rouse1995, saira2020, Wims19a} to be
$10^{-9} H$. Noting that a lower $L$ results in a more harmonic potential during
computation, we set a minimum $L$ to be $0.3 nH$. This is in order to stay
within the parameter range for which $ \beta > 1$ and we can still use the
analytic expressions derived above. To assure $\gamma>\beta$, $\gamma$ values
were tested in the range $[3.0,20.0]$.

After choosing a pair of circuit parameters $L$ and $\gamma$, we turn to
simulation. First, $\VS$ must be chosen by setting $\pxS$ and $\pxdcS$. This is
done by calculating $\pxdcS \equiv \pxdc^{c} + \Delta S$, where
$\pxdc^c(\gamma,\beta)$ is from Eq. (\ref{eq:phi^c_xdc}). The parameter $\Delta
S$ is initialized manually to a value $\Delta S^*$ when starting a new round of
simulations. ($\Delta S^* = 0.16$ was used in the heatmaps shown in Fig.
\ref{fig:heatmaps}.) Then, using the ``min of max'' method (Sec.
\ref{sec:dbneq0}), we set $\pxS$.

Finally, $\VS$ is tested by sampling \ntest\ states from $\VS$'s equilibrium
distribution using a Monte Carlo algorithm. The resulting ensemble is verified
by determining that it contains two well-separated informational states by
asserting that:
\begin{align}
\langle \p < 0 \rangle + 3  \sigma_{\p<0} < \langle \p > 0 \rangle - 3  \sigma_{\p>0}
   ~,
 \end{align}
where $\langle s \rangle$ and $\sigma_s$ are means and standard deviations of
$\p$ conditioned on $s$ being true. If the ensemble fails the test, $\Delta S$
is incremented and the process is repeated. If the ensemble succeeds, we have
found a viable $\VS$.

\begin{figure*}[t]
\centering
\includegraphics[width=\textwidth, trim={0 0 0 0},clip]{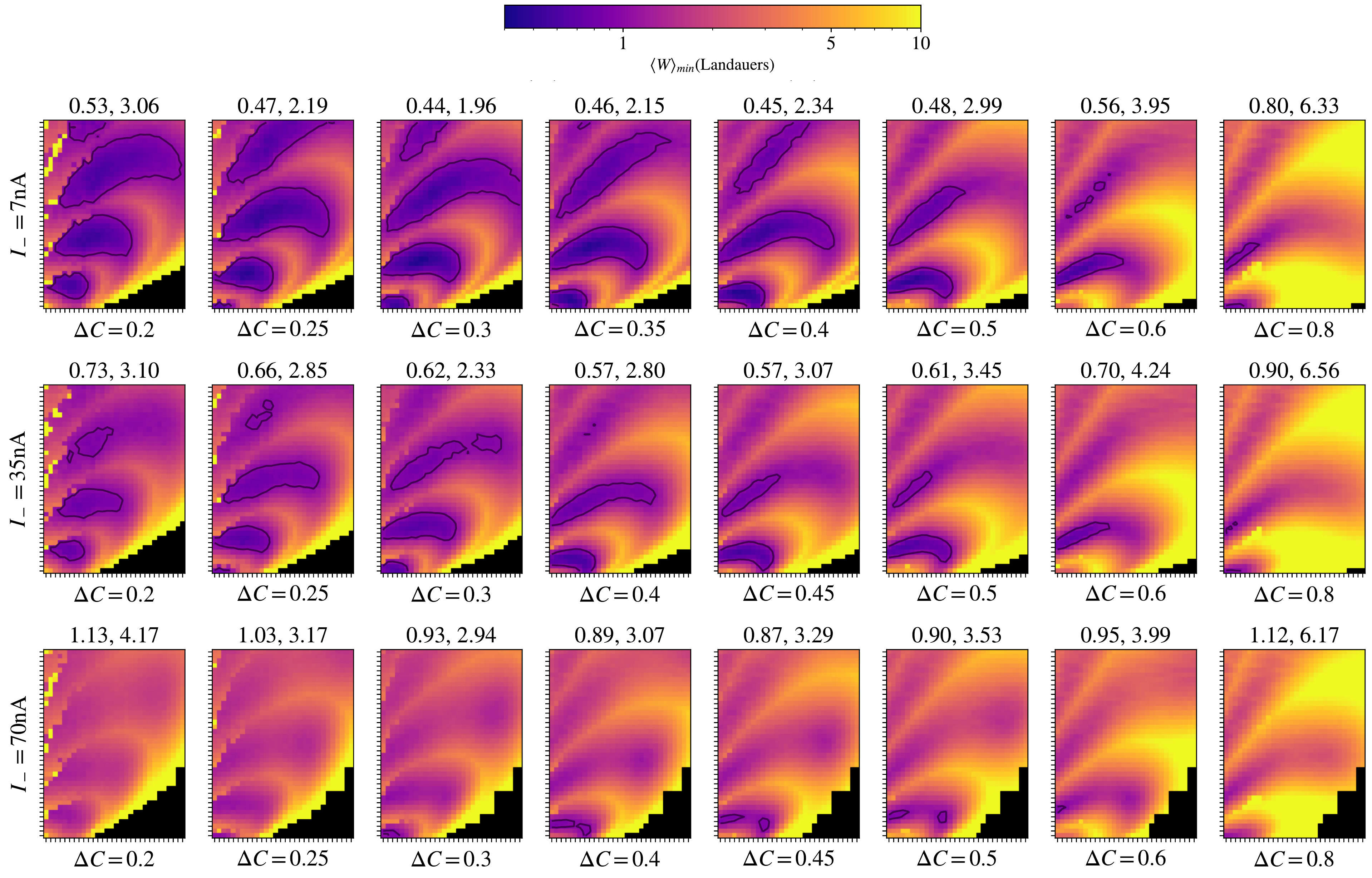}
\caption{Thermodynamic performance under changing $\Delta C$ for devices with
	three different symmetry parameters: In each case, the $x$ axis variable is
	$L\in (0.3,1)$nH and the $y$ axis $\gamma \in (3,20)$. The numerical figures
	at the top of each panel are the minimum and average values of $\langle W
	\rangle_\text{min}$. The outlined (black line) regions represent pieces of
	parameter space where the minimal work protocols cost less than one
	Landauer. The simulations represented by each point in the heatmaps used
	$10,000$ samples from the equilibrium distribution. And, $1,200$ parameter
	sets were tested in each map.
	}
\label{fig:DeltaC}
\end{figure*}

Then, we move on to establish $\VC$ by choosing $\pxC$ and $\pxdcC$. Similar to
$\pxdcS$, $\pxdcC \equiv \pxdc^c-\Delta C$ with $\Delta C$ manually set. The
value of $\Delta C$ does effect the eventual work cost, but the work costs vary
smoothly, and a single value of $\Delta C$ tends to work well over a large
parameter range. Manually setting a single value for $\Delta C$, rather than
allowing it to adjust itself to fall into a local minimum, substantially
reduces simulation run time. However, we expect that given more compute
resources a wider range of sub-Landauer protocols will be discovered. Figure
\ref{fig:DeltaC} shows the effect of changing $\Delta C$ for three different
devices. Once $\Delta C$ is chosen, we use the ``min of mid'' (Sec.
\ref{sec:dbneq0}) method to set $\pxC$ and fully determine $\VC$.

Next, a preliminary simulation is run to identify an approximate value of
the computation time $\tau$. To make the simulation run quickly, the ensemble
above is coarse-grained into two partitions based on whether $\p>0$ or $\p<0$.
Then, each partition is coarse-grained again into $\approx 250$ representative
points through histogramming. A Langevin simulation is run over the histogram
data, exposing it to $\VC$ for a time $\mathcal{O}(10) \sqrt{LC}$. This ensures
capturing the time with the best bit swap. Next, weighting the simulation
results by histogram counts within each partition, we obtain conditional
averages for an approximation of the behavior over the entire ensemble. These
averages are parsed for a set of times at which there are indications of a
successful and low-cost bit swap: $\langle \p(t=0) < 0 \rangle >0$, $\langle
\p(t=0) > 0 \rangle <0$, and values of $\langle \dot{\p} \rangle$ and $\langle
\dot{\pdc} \rangle$ that are close to zero. See, for example, the blue
highlighted portion on the top panel of Fig. \ref{fig:tau_sweep}. In this way,
a range $(\tau_{\min},\tau_{max})$ is determined for $\tau$.

Now, a larger simulation is completed to determine $\tau$ that give the lowest
work value. Another \ntrial\ samples are generated from $\VS$'s equilibrium
distribution, and a Langevin simulation is run on the full ensemble by exposing
it to $\VC$ for $\tau_{max}$ time units. Since the potential is held constant between $t=0$ and $t=\tau$,
work is only done when turning $\VC$ on at $t=0$ and turning it off at
$t=\tau$. The average work done at $t=0$ is $W_0 \equiv \langle
\VC(\p(0),\pdc(0)) - \VS(\p(0),\pdc(0)) \rangle $ and returning to $\VC$ at
time $t$ costs $W_t \equiv \big\langle \VS(\p(t),\pdc(t)) -
\VC(\p(t),\pdc(t))\big\rangle$. Thus, the mean net work cost at time
$t$ is the sum $W(t)=W_0 + W_t$.

Additionally, for each $t \in (\tau_{\min},\tau_{max})$ we calculate the
fidelity $f(t)$ and whether the final states are well-separated informational
states, $s(t)$:
\begin{align}
\begin{split}
f(t) & =
 1- \frac{1}{N} \sum_{i=1}^N \text{bool}\left[\text{sign} \p_i(t=0) = \text{sign}\p_i(t=t)\right] \\
s(t) & =
\text{bool}\left[\langle \p < 0 \rangle + 3  \sigma_{\p<0} < \langle \p > 0 \rangle - 3  \sigma_{\p>0}\right]
~.
\end{split}
\end{align}

Finally, we choose the minimum work protocol via $\text{inf} \left( W(t) :
f(t)\geq 0.99, s(t)=\text{True} \right)$.

After this, we move on to the next pair of $L$ and $\gamma$. Typically, these
are chosen to be individually close to the last pair. And, and instead of
re-initializing $\Delta S$ to its initial value by hand, we decrement $\Delta
S$ from its current value by a small amount if $\Delta S > \Delta S^*$, using
this value as the starting point for the next $L$ and $\gamma$ pair. This
allows the value of $\Delta S$ to drift from its starting point towards more
favorable values as the parameters change, while still preferring to be close
to the known well-behaved parameter value $\Delta S^*$. Setting a new initial
value for $\Delta S$ goes full circle, to find the next minimum work protocol
by repeating the procedure.

This procedure yielded rather large ranges of parameter space over which we
found very low work-cost bit swap protocols. Here, we offer no proof that the
protocols found achieve the global minimum work, since the protocol space is
high dimensional and contains many local minima. That said, improved algorithms and a larger
parameter-range search should result in even lower work costs.

Langevin simulations of the dimensionless equations of motion employed a
fourth-order Runge-Kutta method for the deterministic portion and Euler's
method for the stochastic portion of the integration with $dt$ set to \dt.
(Python NumPy's Gaussian number generator was used to generate the memoryless
Gaussian variable r(t).)

\end{document}